\documentclass[%
reprint,
superscriptaddress,
amsmath,amssymb,
aps,
floatfix,
ekeyshown,
twocolumn
]{revtex4}

\usepackage{color}
\usepackage{comment}
\usepackage{graphicx}
\usepackage{dcolumn}
\usepackage{bm}
\usepackage{multirow}

\usepackage{verbatim}
\usepackage[colorlinks=true,linkcolor=red,urlcolor=blue,citecolor=blue]{hyperref}
\usepackage{soul, xcolor}

\soulregister\cite7
\soulregister\ref7
\soulregister\eqref7

\newcommand{\bchi}{\bar{\chi}}
\newcommand{\hchi}{\hat{\chi}}
\newcommand{\bhchi}{\bar{\hchi}}
\newcommand{\cA}{\mathcal{A}}

\newcommand{\cR}{\mathcal{R}}

\newcommand{\bfF}{\bar{\fF}}
\newcommand{\fF}{\mathsf{F}}
\newcommand{\bOm}{\bar{\Omega}}
\newcommand{\be}{\begin{eqnarray}}
\newcommand{\ee}{\end{eqnarray}}

\newcommand{\tomega}{\kappa}
\newcommand{\absangparam}{\ell}
\newcommand{\angparam}{m_\ell}
\newcommand{\azimuthal}{m_j}
\newcommand{\pmass}{m}
\newcommand{\pcharge}{q}
\newcommand{\dbar}{\lower0.1ex\hbox{$\mathchar'26$}\mkern-12mu d}

\newcommand{\cB}{{\mathcal B}}

\newcommand{\bD}{ \boldsymbol{D}}

\newcommand{\bG}{ \boldsymbol{G}}
\newcommand{\tbG}{{\tilde{\boldsymbol{G}}}}

\newcommand{\bV}{ \boldsymbol{V}}

\newcommand{\cS}{{\mathcal S}}

\newcommand{\dd}{{\mathrm d}}
\newcommand{\cO}{{\mathcal O}}
\newcommand{\bU}{{\boldsymbol{U}}}
\newcommand{\bfM}{{\boldsymbol{M}}}
\newcommand{\bR}{{\boldsymbol{T}}}
\newcommand{\balpha}{{\boldsymbol{\alpha}}}
\newcommand{\re}{{\text{Re }}}
\newcommand{\im}{{\text{Im}\,}}
\newcommand{\ing}{{\text{in}}}
\newcommand{\outg}{{\text{out}}}
\newcommand{\gr}{{\text{up}}}
\newcommand{\de}{{\text{dw}}}

\begin{document}

\title{Quasibound states and late-time evolution of a massive fermion around a Reissner-Nordstr\"{o}m black hole}

\author{Guang-Shang Chen}
\affiliation{Key Laboratory of Particle Physics and Particle Irradiation (MOE), Institute of Frontier and Interdisciplinary Science, Shandong University, Qingdao, 266237, Shandong, China}
\affiliation{CAS Key Laboratory of Theoretical Physics, Institute of Theoretical Physics, 
Chinese Academy of Sciences, Beijing 100190, China}
\affiliation{School of Physical Sciences, University of Chinese Academy of Sciences, No. 19A Yuquan Road, Beijing 100049, China}

\author{Cheng-Bo Yang}
\affiliation{CAS Key Laboratory of Theoretical Physics, Institute of Theoretical Physics, 
Chinese Academy of Sciences, Beijing 100190, China}
\affiliation{School of Physical Sciences, University of Chinese Academy of Sciences, No. 19A Yuquan Road, Beijing 100049, China}

\author{Shou-Shan Bao}
\email[]{ssbao@sdu.edu.cn}
\affiliation{Key Laboratory of Particle Physics and Particle Irradiation (MOE), Institute of Frontier and Interdisciplinary Science, Shandong University, Qingdao, 266237, Shandong, China}

\author{Yue-Liang Wu}
\email[]{ylwu@itp.ac.cn}
\affiliation{CAS Key Laboratory of Theoretical Physics, Institute of Theoretical Physics, 
Chinese Academy of Sciences, Beijing 100190, China}
\affiliation{International Centre for Theoretical Physics Asia-Pacific (ICTP-AP), UCAS, Beijing 100190, China}
\affiliation{Taiji Laboratory for Gravitational Wave Universe (Beijing/Hangzhou), University of Chinese Academy of Sciences (UCAS), Beijing 100049, China}
\affiliation{School of Fundamental Physics and Mathematical Sciences, Hangzhou Institute for Advanced Study, UCAS, Hangzhou 310024, China}

\author{Hong Zhang}
\email[]{hong.zhang@sdu.edu.cn}
\affiliation{Key Laboratory of Particle Physics and Particle Irradiation (MOE), Institute of Frontier and Interdisciplinary Science, Shandong University, Qingdao, 266237, Shandong, China}

\date{\today}

\begin{abstract}
A massive fermion around a charged black hole provides a gravitational analog of atomic bound states and their relaxation.  In this work, we study this system by formulating the radial equation as a coupled matrix system and constructing the Green's function with ingoing boundary conditions at the horizon and decaying boundary conditions at infinity. In the weak-coupling scenario $|qQ|\sim mM<1$, a matrix matching scheme gives an improved analytic expression of quasibound-state spectrum, including fine-structure corrections and more accurate decay widths. The extremal Reissner-Nordstr\"{o}m case ($|Q|=M$) is treated separately and shown to be the smooth limiting result of the nonextremal spectrum. We further analyze the branch-cut contribution to the time-domain Green's function in the late-time limit. We confirm an oscillatory power-law behavior in intermediate late-time regime $1/m < t < 1/m^3M^2$. In the far late-time regime $t>1/m^3M^2$, the activation of the quasibound states produces an $t^{-5/6}\exp(-\eta t^{1/3})$ suppression with a chirping phase before the asymptotic $t^{-5/6}$ tail previously found in the limit $t\to\infty$. Direct time-domain simulations support this distinction and show how the quasibound contribution coexists with the familiar power-law component. 
\end{abstract}

    \maketitle

\section{Introduction}\label{sec:intro}
    
Much as the study of a fermion bound to a nucleus elucidated the quantum era, the gravitational counterpart of the atom is anticipated to serve as a crucial probe for signatures of quantum gravity \cite{Hod:1998vk, Dolan:2015eua}. This fermion-black-hole analog, formed by a Dirac field gravitationally interacting with the black hole, provides a special case to explore the interplay between gravity and quantum field theory. In this work, we would explore the system composed of a charged massive fermion and a Reissner-Nordstr\"{o}m (RN) black hole. 

Generally, the dynamics of matter fields around a black hole are governed by two distinct families of spectral modes with complex eigenfrequencies. The quasinormal modes (QNMs) are radiative and dissipative states outwardly propagating at infinity \cite{Konoplya:2011qq, Berti:2009kk}. In contrast, the quasibound states (QBSs) \cite{GaltsovD:1983zpz,Detweiler:1980uk,Dolan:2015eua,Huang:2017nho}, on which this paper focuses, are characterized by the exponentially decaying behavior far from the horizon. These localized modes are relevant only for massive fields and were also referred to  as "graviatoms" \cite{Laptev:2006xz, Arvanitaki:2010sy} or "quasilevels" \cite{Gaina:1992nx} in the literature.

The dynamics of bosonic (including scalar, vector and tensor) QBSs has been studied extensively, yielding a deep understanding of their spectra \cite{Detweiler:1980uk, Leaver:1990zz, Rosa:2011my, Dolan:2007mj, Rosa:2011my, Bao:2022hew, Bao:2023xna, Chu:2024iie}, wigs \cite{Barranco:2012qs, Sanchis-Gual:2017bhw}, the superradiance instability \cite{Brito:2015oca, Cardoso:2017kgn, Barack:2018yly, Cardoso:2018nvb, Cardoso:2018tly, Spieksma:2023vwl, Dias:2023ynv, Spieksma:2024voy, Berti:2025hly} and possible observables \cite{Chen:2019fsq, Chen:2022nbb, Spieksma:2023vwl, Chen:2022kzv, Guo:2022mpr, Guo:2024dqd, Jia:2025vqn, Guo:2025dkx}. However, fermionic fields remain relatively unexplored, largely due to the absence of superradiance \cite{Gueven:1977dq, Martellini:1977qf, Lee:1977gk, Iyer:1978du}. Building on the foundational works on the separability of Dirac equation \cite{Unruh:1973bda, Chandrasekhar:1976ap, Unruh:1976fm, Carter:1979fe}, fermionic QBS spectra were initially calculated using subsequent matching technique and nonrelativistic approximation \cite{Ternov:1980st, Gaina:1988nf}. These works formulate the problem in terms of a decoupled Klein-Gordon-like equation, which is technically convenient but may obscure the original first-order structure and affect the control of subleading terms. We therefore adopt the coupled first-order formulation, where the two radial components and their boundary conditions are treated on equal footing.  In view of the substantial differences \cite{Aretakis:2011ha,Aretakis:2011hc,Lucietti:2012sf} between extremal and nonextremal Reissner-Nordström black holes, the extremal case is independently analyzed in this work.

Beyond these analytical approximations, the spectra of QBSs have been numerically determined. A widely used benchmark is provided by improved continued-fraction techniques \cite{Rosa:2011my, Dolan:2015eua, Huang:2017nho, Konoplya:2017tvu, Jing:2005dt} which were originally proposed for scalar QNMs \cite{Leaver:1985ax, Nollert:1993zz}. Our newly proposed expression for nonextremal RN will be validated by comparing it with numerical results obtained from a matrix-valued implementation of this approach. However, the application of this scheme to fields around an extremal black hole involves additional subtleties. An improved analytic expression for the QBS spectra is derived using the recently developed matrix matching method \cite{Chen:2025enc}. To further validate our results, we compare them with those obtained from an independent shooting method \cite{Lasenby:2002mc, Giammatteo:2004wp, Dolan:2009kj}. 

Strictly speaking, however, a description relying solely on discrete spectra yields an incomplete picture of the field dynamics. The dissipative boundary condition at the event horizon leads to the non-Hermitian nature of the time-evolution operator, implying that the pole expansion alone does not, in general, give a complete representation of the retarded response; branch-cut contributions must also be included \cite{Leaver:1986gd, Nollert:1999ji, Andersson:1996cm, Ching:1994bd, Casals:2013mpa}. A self-consistent description necessitates the inclusion of contribution from branch cuts in the complex frequency plane \cite{Gundlach:1993tp}. Alongside the pole contributions, the branch-cut contribution dominates a power-law tail at sufficiently late times and forms distinctive behavior at asymptotic late time \cite{Price:1971fb, Hod:1998ra, Koyama:2001ee, Jing:2004xv, Jing:2005uy}, critically governing the ultimate relaxation of the field. 

Within the same matrix-formulation framework, we analyze the late-time behavior of fermionic QBS by explicitly incorporating boundary conditions at both the event horizon and spatial infinity in Green's function. This contrasts with the majority of the existing studies primarily investigating the ultimate relaxation of the perturbed spacetime. Our results confirm the role of long-lived QBS activation in the late-time dynamics, which is supported by numerical verification  up to $t=5\times 10^5$, using direct time-domain simulation.

The paper is organized as follows. We begin with an introduction of our master formalism in Sec.~\ref{sec:formalism}. Besides the tetrad formalism and the Newman-Penrose formalism widely adopted in the literature, we reobtain the equation based on gravitational quantum field theory. A thorough analysis of the QBS spectra in the weak-coupling scenario is presented in Sec.~\ref{sec:spectrum}. We distinguish the extremal and nonextremal black hole and treat them separately. In Sec.~\ref{sec:latetime}, the contribution from the branch cuts is calculated explicitly and simulated numerically, which yields a consistent result. Finally, concluding remarks are given in Sec.~\ref{sec:summary}.

\section{Formalism}\label{sec:formalism}

\subsection{Dirac equation around a RN black hole}

Based on the fundamental premise that the laws of nature are governed by intrinsic properties of the basic constituents of matter, the \textit{gravitational quantum field theory} and the subsequent 
\textit{generalized standard model} establish a comprehensive framework to reconcile general relativity (GR) and quantum field theory \cite{Wu:2015wwa, Wu:2017rmd, Wu:2017urh, Wu:2022mzr, Wu:2022aet, Wu:2024mul, Wu:2025abi, Wu:2025rei, Gao:2024juf, Gao:2025aye, Xu:2025yrn}. Within this framework, intrinsic and external symmetries are rigorously distinguished. The global Lorentz symmetry SO(1, 3) is defined in Minkowski spacetime where fields like spinors live. Concurrently, the intrinsic spin symmetry SP(1, 3) is defined in Hilbert space and localized in accordance with the gauge principle. These two are unified as joint symmetries SO(1, 3)$\Join$SP(1, 3), mediated by a spin-related vector field $\hat\chi^\mu_a(x)$ that transforms homogeneously under both groups.

In the theory, the gravigauge field $\chi^a_\mu$ is defined as the inverse of $\hchi^\mu_a$. Endowed simultaneously with coordinate index (\textit{Greek}) and spin index (\textit{Latin}), it is identified as the fundamental degree of freedom of gravity and distinguished from the metric field $g_{\mu\nu}$ in GR. Mathematically, $\chi^a_\mu$ is understood as a gauge-type field sided in Minkowski spacetime and valued in the spin-related gravigauge spacetime $\mathbf G_{4}$, which conceptualizes a biframe spacetime and a fiber bundle with the spin-related intrinsic gravigauge spacetime as its fiber \cite{Wu:2015wwa, Wu:2017rmd, Wu:2017urh, Wu:2022mzr, Wu:2022aet, Wu:2024mul, Wu:2025abi, Wu:2025rei}. Incorporating with $\chi^a_\mu$ and the spin gauge field $\cA^{ab}_\mu$, a spin-gauge-invariant action can be uniquely fixed. 

Explicitly, a fermion governed by electromagnetic, gravitational and spin-related interactions $\Psi$ obeys the generalized Dirac equation:
\begin{equation}\label{eq:generalized_Dirac}
	\gamma^a\hat\chi_{a}^{\; \mu}\left(i\partial_{\mu} - iV_{\mu} - \pcharge A_\mu + \mathcal A_{\mu}\right)\Psi =\pmass\Psi,
\end{equation}
where $\gamma^a$ are gamma matrices defined in flat spacetime and in the Dirac representation. Notations $\pmass$ and $\pcharge$ are correspondingly the mass and the charge of the field. $A_\mu$ is referred to as electromagnetic U(1) gauge field. The spin gauge field governed by SP(1, 3) is defined as
\begin{equation}
	\mathcal A_{\mu} = \mathcal A_{\mu}^{ab}\frac12\Sigma_{ab}, \quad \Sigma_{ab}  = \frac{i}{4}[\gamma_a, \gamma_b].
\end{equation}
Furthermore, $V_\mu$ is an induced vector field that ensures the Weyl symmetry and is in the following form:
\begin{align}
    V_{\mu} &\equiv \tfrac{1}{2} \chi \, \hat{\chi}_{b}^{\;\; \nu}\mathcal{D}_{\nu}(\hat{\chi} \chi_{\mu}^{\;\; b}), \\
    \mathcal{D}_{\nu}(\hat{\chi}\chi_{\mu}^{\;\;b}) &= \partial_{\nu} (\hat{\chi}\chi_{\mu}^{\;\; b}) + \hat{\chi} \mathcal{A}_{\nu\, c}^{b}  \chi_{\mu}^{\;\;c}.
\end{align}

For analytical convenience, the generalized Dirac equation can be recast into an equivalent form:
\begin{equation}
    \gamma^ai\hat\chi_{a}^{\; \mu}\left(\partial_{\mu} + \hat{V}_{\mu} - i\hat{A}_{\mu}\gamma^5 + iqA_\mu\right)\Psi = \pmass\Psi,
\end{equation}
with the vector and axial vector fields
\begin{align}
    \hat{V}_{\mu} &\equiv \tfrac{1}{2} \hat{\chi}_{b}^{\; \nu} \mathcal{A}_{\nu c}^{b}\chi_{\mu}^{\; c} - V_{\mu}, \\
    \hat{A}_{\mu} &\equiv \tfrac{1}{4} \epsilon_{cdc'd'} \chi_{\mu}^{\; c} \hat{\chi}^{\nu d} \mathcal{A}_{\nu}^{c'd'}. 
\end{align}
In this work, we focus on the regime where the internal contribution of $\cA^{ab}_\mu$ and any effects violating the equivalence principle are neglected. As shown in our previous analysis \cite{Chen:2025enc}, the spin gauge field is treated as a background field identified with the spin connection:
\begin{equation}
    \begin{aligned}
        \langle \cA_{\mu}^{ab}(x) \rangle &\equiv \bOm_{\mu}^{ab}(x) \\
        &= \tfrac{1}{2}\left( \bhchi^{\nu a} \bfF_{\mu\nu}^{b} - \bhchi^{\nu b} \bfF_{\mu\nu}^{a} -  \bhchi^{\rho a}  \bhchi^{\sigma b}  \bfF_{\rho\sigma}^{c} \bchi_{\mu c } \right).
    \end{aligned}
\end{equation}
Consequently, it ceases to be an independent degree of freedom in the theory and is entirely determined by the background gravigauge field $\bchi^a_\mu$, where 
\begin{align}
    \bfF_{\mu\nu}^{a} =\partial_{\mu}\bchi^a_{\nu} - \partial_{\nu}\bchi^a_{\mu}.    
\end{align}
Correspondingly, the vector and axial vector fields are simplified to:
\begin{equation}
    \begin{aligned}
        \langle \hat{V}_{\mu} \rangle &= \tfrac{1}{2} \bhchi_{b}^{\; \nu} \bOm_{\nu c}^{b} \bchi_{\mu}^{\; c}, \\
        \langle \hat{A}_{\mu} \rangle &= \tfrac{1}{4} \epsilon_{cdc'd'} \bchi_{\mu}^{\; c} \bhchi^{\nu d} \bOm_{\nu}^{c'd'}.
    \end{aligned}
\end{equation}
Under the background field approximation, the full theory recovers GR as its classical limit. The metric of the classical curved spacetime is thus induced by the background gravigauge field
\begin{equation}\label{eq:metric2gravigauge}
    g_{\mu\nu} \equiv \bchi_{\mu\nu} = \eta_{ab}\bchi^a_\mu\bchi^b_\nu.
\end{equation}

Such a formalism enables us to deduce the background gravigauge fields from classical theories, circumventing intricate nonlinear equations in the full theory. Classically, a static, charged and spherically symmetric black hole is described by the Reissner-Nordstr\"{o}m metric 
\begin{equation}
    g_{\mu\nu} = \frac{\Delta(r)}{r^2}\dd t^2  - \frac{r^2}{\Delta(r)}\dd r^2 - r^2(\dd\theta^2 + \sin^2\theta\dd\varphi^2).
\end{equation}
Here $\Delta(r)$ is a quadratic polynomial defined as
\begin{equation}
    \Delta(r) = r^2-2Mr+Q^2 = (r-r_-)(r-r_+),
\end{equation}
where $M$ and $Q$ denote the mass and charge, respectively, of the black hole. The parameters $r_\pm$ are the roots of the polynomial satisfying $0<r_-\leq r_+$. The corresponding electromagnetic gauge field is set to be
\begin{equation}
    A_0 = \frac{Q}{r}
\end{equation}
with the other three components vanishing after gauge fixing. Throughout this article, we adopt the Planck units $G=\hbar=c=1$ for convenience. For the case $Q<M$, the roots $r_\pm$ of $\Delta(r)$ are real and correspond, respectively, to the Cauchy horizon ($r_-$) and the event horizon ($r_+$) of the black hole. In the special case $|Q|=M$, the two horizons coincide and the black hole is called extremal Reissner-Nordstr\"{o}m (eXRN).

In the exterior region $r>r_+$, the RN metric corresponds to the following background gravigauge field according to Eq.~\eqref{eq:metric2gravigauge}:
\begin{equation}
    \begin{aligned}
        \bchi^0_t &= \frac{\sqrt{\Delta(r)}}{r}, &
        \bchi^1_r &= \frac{r}{\sqrt{\Delta(r)}}, \\
        \bchi^2_\theta &= r, &
        \bchi^3_\varphi &= r\sin\theta. \\
    \end{aligned}
\end{equation}
All other 12 unlisted components vanish identically. Hence, the generalized Dirac equation assumes the following explicit form:
\begin{equation}\label{eq:Dirac_RN}
    \bigg[\frac{\gamma^0\sqrt{\Delta(r)}}{r}iD_t
    + \frac{\gamma^1r}{\sqrt{\Delta(r)}}iD_r
    + \frac{i\gamma^1\gamma^0\hat K}{r} - \pmass\bigg]\Psi = 0.
\end{equation}
Here, we have also introduced the following notations: 
\begin{align}
    D_t &\equiv \partial_t + \frac{i\pcharge Q} {r}, \\
    D_r &\equiv \partial_r + \frac1{2r} + \frac1{4(r-r_-)} + \frac1{4(r-r_+)}.
\end{align}
The so-called $K$ operator is defined on the two-sphere $S^2$. In our choice of representation, it is explicitly shown as follows:
\begin{equation}
    \hat K = i\gamma^0\left[\gamma^1\gamma^2i\left(\partial_\theta + \frac{\cot\theta}2\right) + \frac1{\sin\theta}\gamma^1\gamma
    ^3i\partial_\varphi\right].
\end{equation}
Such an operator admits two linearly independent eigenstates $\Phi^\pm_{\angparam,\azimuthal}(\theta,\varphi)$ \cite{Chen:2025enc}, labeled by two numbers
\begin{align}
    \angparam &= \pm\absangparam, \\
    \azimuthal &= \pm\frac12, \pm\frac32,\dots,\pm\absangparam-\frac12,
\end{align}
for arbitrary positive integer $\absangparam>0$. 

Accordingly, the spinor field $\Psi$ in Eq.~\eqref{eq:Dirac_RN} can be decomposed as 
\begin{equation}
    \begin{aligned}
        \Psi = \frac{\psi^+_{\angparam}(t,r)\Phi^+_{\angparam,\azimuthal}(\theta,\varphi) +\psi^-_{\angparam}(t,r)\Phi^-_{\angparam,\azimuthal}(\theta,\varphi)}{r^{1/2}\Delta^{1/4}(r)},
    \end{aligned}
\end{equation}
where the radial functions $\psi^\pm_{\angparam}(t,r)$ satisfy the following coupled partial differential equations:
\begin{equation}\label{eq:radial}
    \begin{aligned}
        \bigg(\frac{\sqrt{\Delta(r)}}{r}\partial_r + \frac{r}{\sqrt{\Delta(r)}}\partial_t + \frac{i\pcharge Q}{\sqrt{\Delta(r)}}\bigg)\psi^+_{\angparam}(t,r) & \\
        \ = -\left(\frac{\angparam}{r} - i\pmass\right)\psi^-_{\angparam}(t,r) &, \\
        \bigg(\frac{\sqrt{\Delta(r)}}{r}\partial_r - \frac{r}{\sqrt{\Delta(r)}}\partial_t - \frac{i\pcharge Q}{\sqrt{\Delta(r)}}\bigg)\psi^-_{\angparam}(t,r) & \\
        \ = -\left(\frac{\angparam}{r} + i\pmass\right)\psi^+_{\angparam}(t,r) &. \\
    \end{aligned}
\end{equation}
This radial system is consistent with the equation presented in the literature and derived from the tetrad formalism or the Newman-Penrose formalism. The remainder of this article will focus exclusively on these equations for fixed values of $\azimuthal$ and $\angparam$. When no ambiguity arises, subscripts will be omitted in the following discussion for convenience.

\subsection{Radial Green's functions in time and frequency domain}

To facilitate the analysis, we apply several transformations to the master equation. We first introduce the dimensionless variables $x=r/M-1$ and the Fourier-transformed field $\psi^\pm(\omega,x)$ defined in the frequency domain. By introducing a two-component field
\begin{equation}
    \psi(\omega,x) = \begin{pmatrix}\psi^+(\omega,x) \\ \psi^-(\omega,x)\end{pmatrix},
\end{equation}
Eq.~\eqref{eq:radial} is recast into a compact operator form: 
\begin{equation}\label{eq:radial_matrix}
    \bD(x)\psi(\omega, x) = 0.
\end{equation}
The operator $\bD(x)$ and the matrix-valued potential $\bV(x)$ are defined as
\begin{align}
    \bD(x) &= \boldsymbol{I}\partial_x - \frac{(x+1)^2}{x^2-b^2}\bigg[i\omega M\boldsymbol{\sigma}_3 - \bV(x)\bigg], \\
    \bV(x) &= \dfrac{i\pcharge Q\boldsymbol{\sigma}_3}{x+1} + \dfrac{\angparam\sqrt{x^2-b^2}}{(x+1)^2}\boldsymbol{\sigma}_1 + \frac{\pmass M\sqrt{x^2
    -b^2}}{x+1}\boldsymbol{\sigma}_2, 
\end{align}
where $\boldsymbol{I}$ is the identity matrix, $\bV(x)$ is a combination of Pauli sigma matrices $\boldsymbol{\sigma}_{1,2,3}$ and we have the algebraic relationship $\bV^T = (i\boldsymbol{\sigma_2})\bV(i\boldsymbol{\sigma_2})$. At the same time, a dimensionless parameter 
\begin{equation}
    b = \frac{r_+-r_-}{r_++r_-}
\end{equation}
is introduced to measure the deviation from extremality of the hole. Specifically, $b=0$ corresponds to the eXRN black hole and we recover the Schwarzschild black hole for $b=1$.

Given an initial configuration at $t=0$, the subsequent evolution of the field is completely described by a matrix-valued Green's function $\bG(x,y;t)$. The time-domain Green's function is related to its frequency-domain counterpart via the transform:
\begin{equation}\label{eq:temporal_green}
    \bG(x,y;t) = \frac1{2\pi}\int_{\mathbb{R}+i\epsilon} \dd\omega\ e^{-i\omega t}\tbG(x,y;\omega).
\end{equation}
Within the integrand, $\tbG$ is the solution to the inhomogeneous differential equation
\begin{equation}
    \bD(x)\tbG(x,y;\omega) = \delta(x-y)\boldsymbol{I},
\end{equation}
corresponding to Eq.~\eqref{eq:radial_matrix}. Generally, Eq.~\eqref{eq:radial_matrix} admits two linearly independent solutions $\psi_\ing$ and $\psi_\outg$, where $\psi_{\ing}$ ($\psi_{\outg}$) denotes asymptotically pure ingoing (outgoing) waves near the horizon. Alternatively, categorizing by their asymptotic behavior at spatial infinity yields another group of independent solutions $\psi_\de$ and $\psi_\gr$. While the energy is insufficient to surmount the mass barrier, $\psi_\de$ ($\psi_\gr$) represents the exponentially decaying (growing) modes toward spatial infinity, respectively. Conversely, for superbarrier energies, $\psi_\de$ ($\psi_\gr$) corresponds to purely ingoing (outgoing) waves in the far region.

The event horizon of a classical black hole acts as a one-way membrane that permits only inward flux. Such a principle enforces the boundary condition that the field must be pure ingoing at $x=b$. On the other hand, this study focuses on the fermions localized near the black hole. The appropriate far-field boundary condition selects the specific solution $\psi_\de$. This contrasts with the extensive analysis of quasinormal modes, which impose purely outgoing boundary conditions at infinity. These two boundary conditions determine the frequency-domain Green's function: 
\begin{equation}
    \tbG(x,y;\omega) = \begin{cases}
        \dfrac{\psi_\ing(\omega,x)\bar\psi_\de(\omega,y)}{\mathcal W(\psi_\ing,\psi_\de)} & (x<y), \\
        \dfrac{\psi_\de(\omega,x)\bar\psi_\ing(\omega,y)}{\mathcal W(\psi_\ing,\psi_\de)} & (x>y),
    \end{cases}
\end{equation}
where $\bar\psi=i\psi^T\boldsymbol{\sigma}_2$ is the symplectic dual of $\psi$ and the Wronskian 
\begin{equation}
    \mathcal W(\psi_\ing,\psi_\de) = \psi^+_\ing\psi^-_\de - \psi_\de^+\psi^-_\ing.
\end{equation}
It is straightforward to verify from Eq.~\eqref{eq:radial_matrix} that the Wronskian is position independent, namely $\mathrm d\mathcal W/\mathrm dx = 0$.

\begin{figure}
    \centering
    \includegraphics[width=0.9\linewidth]{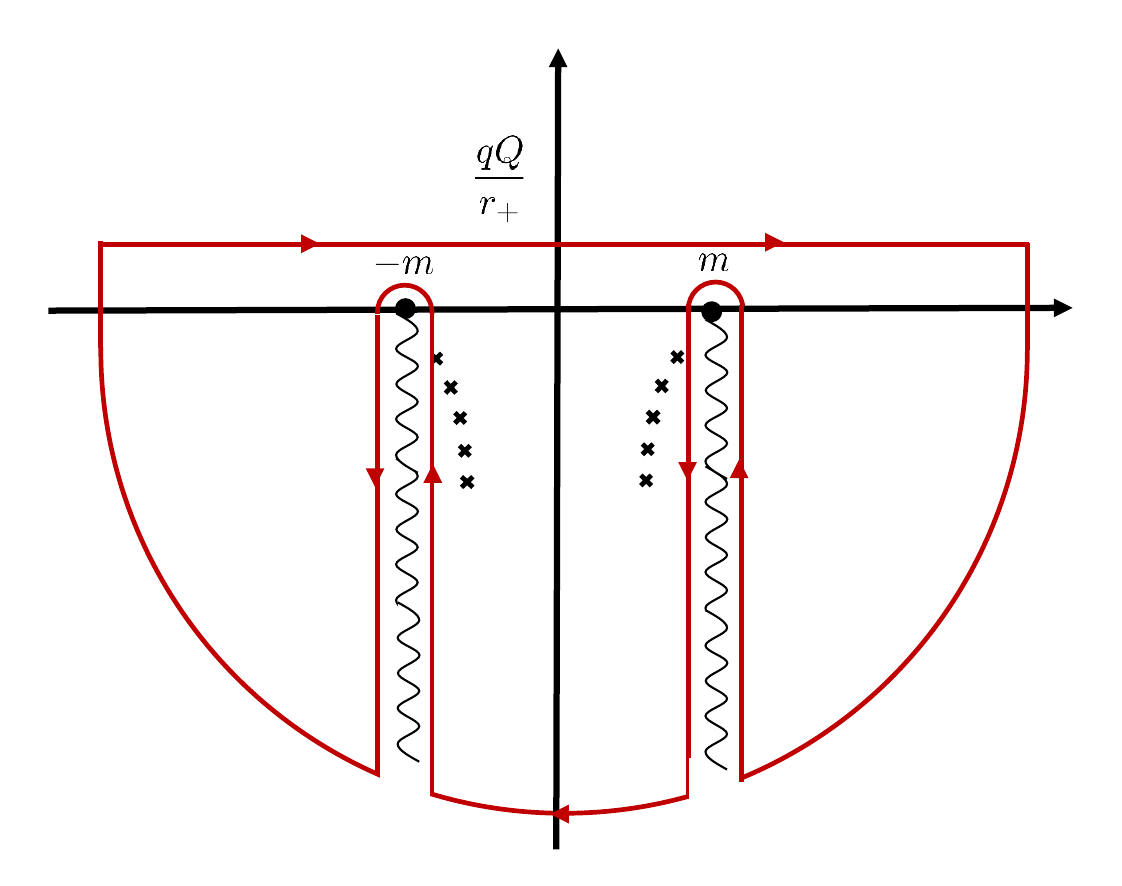}
    \caption{Illustration of the contour on the complex $\omega$ plane. Wavy lines denote branch cuts, and crosses indicate poles. The red directed line represents the closed contour along which the integral is defined. } 
    \label{fig:ComplexPath}
\end{figure}

The concrete time evolution of the field relies on a detailed calculation of the integral in Eq.~\eqref{eq:temporal_green}. A common strategy is to treat it as a contour integral in the complex $\omega$ plane. The field dynamics are then completely determined by its analytic structure, primarily governed by its poles and the branch cuts shown in Fig.~\ref{fig:ComplexPath}. With the contour illustrated by the red curve, the integral subsequently decomposes the retarded Green's function into three parts:
\begin{equation}
    \bG = \bG_{\text{pole}} + \bG_{\text{cut}} + \bG_{\text{arc}}.
\end{equation}
Here, the first term captures the contribution from QBS poles. The second term $\boldsymbol{G}_{\text{cut}}$ arises from the branch cuts associated with the asymptotic momentum
\begin{equation}\label{eq:def_p}
    p=M\sqrt{m^2-\omega^2}
\end{equation}
that is crucial to defining $\psi_\de$ and $\psi_\gr$. The square-root function is multivalued. In this work, we have limited the contour on the Riemann sheet with $\re p>0$. The last term $\boldsymbol{G}_{\text{arc}}$, corresponding to the large semicircle in the lower half-plane, characterizes prompt disturbances that decay away. 

\section{The quasibound-state spectrum}\label{sec:spectrum}

\subsection{Approximate solutions outside nonextremal Reissner-Nordstr\"{o}m black hole}
\label{sec:approxsols}

The nontrivial structure of the potential $\boldsymbol{V}(x)$ makes a full analytical treatment of Eq.~\eqref{eq:radial_matrix} intractable. Therefore, we turn to adopt an approximation scheme by solving it in the neighborhood of the event horizon $b<x<\absangparam/\pmass M$ and in the far-field limit $x> \sqrt{\absangparam/\pmass M}$. Details about the determination of these ranges can be found in Appendix~\ref{app:valid_range}. In the scenario
\begin{equation}\label{eq:assumption}
    |\pcharge Q| \sim |\omega M| \sim \pmass M \ll \absangparam,
\end{equation}
the existence of an overlapping region facilitates the matching procedure required to ascertain the physical quantities of interest, as already demonstrated in our previous work \cite{Chen:2025enc}.

In the \textit{far} region $x>\sqrt{\absangparam/\pmass M}$, we keep the terms up to order $\cO(x^{-1})$ in Eq.~\eqref{eq:radial_matrix} and apply the transformation
\begin{equation}\label{eq:psi_tranformation}
    \begin{aligned}
        \tilde\psi_f^+ &= \left(\frac12 + \frac{2i\omega M-i\pcharge Q}{2\tilde\absangparam}\right)\psi_f^- 
        + \frac{\angparam - i\pmass M}{2\tilde\absangparam}\psi_f^+, \\
        \tilde\psi_f^- &= \left(\frac12 - \frac{2i\omega M-i\pcharge Q}{2\tilde\absangparam}\right)\psi_f^- 
        - \frac{\angparam - i\pmass M}{2\tilde\absangparam}\psi_f^+, \\
    \end{aligned}
\end{equation}
which diagonalizes the coefficient associated with the $1/x$ term. It leads to the approximate equations as follows:
\begin{equation}\label{eq:radial_far}
    \begin{aligned}
        \left(\partial_x + \frac{\tilde\absangparam}{x} - \frac{\tomega}{\tilde\absangparam}\right)\tilde\psi^+_f
        &= -i\beta_-\tilde\psi_f^-, \\
        \left(\partial_x - \frac{\tilde\absangparam}{x} + \frac{\tomega}{\tilde\absangparam}\right)\tilde\psi^-_f 
        &= -i\beta_+\tilde\psi_f^+. \\
    \end{aligned}
\end{equation}
Here we introduce the following notations:
\begin{align}
    \label{eq:eff_ell}
    \tilde\absangparam &= \sqrt{\absangparam^2+\pmass^2M^2-(2\omega M-\pcharge Q)^2}, \\
    \beta_\pm &= \frac{\angparam\omega M- i\pmass M(\omega M -\pcharge Q)}{\angparam+i\pmass M}\left(1\pm\frac{2\omega M-\pcharge Q}{i\tilde\absangparam}\right) \nonumber \\
    &\quad\ \pm \pmass M\frac{\angparam}{\tilde\absangparam}, \\
    \label{eq:def_tomega}
    \tomega &= (2\omega^2-\pmass^2)M^2 - \omega M\pcharge Q.
\end{align} 

By decoupling the system in Eq.~\eqref{eq:radial_far}, the components are determined by second-order equations
\begin{equation}
    \left(\partial_x^2 - p^2 - \frac{2\kappa}{x} + \frac{\tilde\absangparam(\tilde\absangparam\pm1)}{x^2}\right)\tilde\psi^{\pm}_f = 0.
\end{equation}
By analogy with the radial equation of atomic hydrogen, the localized states exist only when $\re\kappa>0$. Hence, this implies that the inequality 
\begin{equation}\label{eq:qbs_condition}
    \re[\omega M(\omega M - \pcharge Q)] > 0
\end{equation}
holds. The solution of the approximate equations is presented in terms of Whittaker functions as
\begin{equation}\label{eq:psif}
    \tilde\psi_f \propto \begin{pmatrix}
        \frac{p\tilde\absangparam-\tomega}{i\tilde\absangparam\beta_+}W_{\tomega/p,\tilde\absangparam+\frac12}(2px) \\
        W_{\tomega/p,\tilde\absangparam-\frac12}(2px)
    \end{pmatrix},
\end{equation}
enabling an approximation of $\psi_\de$ satisfying the decaying boundary condition.

On the other hand, Eq.~\eqref{eq:radial_matrix} is approximated in the \textit{near} region $b<x<\absangparam/\pmass M$ by
\begin{equation}\label{eq:radial_near}
    \begin{aligned}
        \left(\sqrt{x^2-b^2}\partial_x - \frac{2i\omega_h b}{\sqrt{x^2-b^2}}\right)\psi_h^+ + s\tilde\absangparam\psi^-_h &= 0, \\
        \left(\sqrt{x^2-b^2}\partial_x + \frac{2i\omega_h b}{\sqrt{x^2-b^2}}\right)\psi_h^- + s\tilde\absangparam\psi^+_h &= 0. \\
    \end{aligned}
\end{equation}
Here, we denote $s=\text{sgn}\angparam$ and the near-horizon effective frequency
\begin{equation}
    \omega_h = \frac{b+1}{2b}(\omega r_+-\pcharge Q).
\end{equation}
The effective frequency diverges at the extremality limit $b\to0$. It thus calls for a separated analysis of the field around an extremal RN black hole, which is left for the next section. Correspondingly, approximate solutions are in terms of hypergeometric functions:
\begin{gather}\label{eq:psih}
    \begin{aligned}
        \psi_{h1} &\approx \left(\frac{x+b}{x-b}\right)^{i\omega_h}\begin{pmatrix}
            \sqrt{x^2-b^2} F^{1+\tilde\absangparam, 1-\tilde\absangparam}_{\frac32-2i\omega_h}(\frac{b-x}{2b}) \\
            \frac{s(2i\omega_h-\frac12) }{\tilde\absangparam}
            F^{\tilde\absangparam, -\tilde\absangparam}_{\frac12-2i\omega_h}(\frac{b-x}{2b})
        \end{pmatrix}, \\
        \psi_{h2} &\approx \left(\frac{x-b}{x+b}\right)^{i\omega_h}\begin{pmatrix}
            \frac{-s(2i\omega_h+\frac12) }{\tilde\absangparam}
            F^{\tilde\absangparam, -\tilde\absangparam}_{\frac12+2i\omega_h}(\frac{b-x}{2b}) \\
            \sqrt{x^2-b^2} F^{1+\tilde\absangparam, 1-\tilde\absangparam}_{\frac32+2i\omega_h}(\frac{b-x}{2b})
        \end{pmatrix}.
    \end{aligned}
\end{gather}
These two functions depend on $\omega_h$ and it is not hard to show that $\psi_{h1}=\boldsymbol{\sigma_1}\psi_{h1}^*$ . When the inequality \eqref{eq:qbs_condition} holds, the ingoing solution is approximated by $\psi_{h1}$. 

In the scenario in Eq.~\eqref{eq:assumption}, the \textit{overlapping} region
\begin{equation}
    \sqrt{\frac{\absangparam}{\pmass M}} < x < \frac{\absangparam}{\pmass M}
\end{equation}
is nonempty, within which both Eqs.~\eqref{eq:radial_far} and \eqref{eq:radial_near} hold simultaneously. Moreover, the solutions are adequately described by the leading term in asymptotic expansion with respect to coordinate $x$. Concretely, we take
\begin{equation}\label{eq:h2f}
    \begin{aligned}
        \psi_{h1} \approx &\frac{(2b)^{\tilde\absangparam}\Gamma(-2\tilde\absangparam)\Gamma(\frac32-2i\omega_h)}{\Gamma(1-\tilde\absangparam)\Gamma(\frac12-\tilde\absangparam-2i\omega_h)}x^{-\tilde\absangparam}
        \begin{pmatrix} -1 \\ s \end{pmatrix}\\
        &-\frac{(2b)^{-\tilde\absangparam}\Gamma(2\tilde\absangparam)\Gamma(\frac32-2i\omega_h)}{\Gamma(1+\tilde\absangparam)\Gamma(\frac12+\tilde\absangparam-2i\omega_h)} x^{+\tilde\absangparam}
        \begin{pmatrix} 1 \\ s \end{pmatrix}.
    \end{aligned}
\end{equation}
At the same time, it is shown that 
\begin{equation}\label{eq:f2h}
    \begin{aligned}
        \psi_f &\approx 
        \frac{(2p)^{-\tilde\absangparam+1}\Gamma(2\tilde\absangparam)}{i\beta_+\Gamma(\tilde\absangparam-\tomega/p)} x^{-\tilde\absangparam}
        \begin{pmatrix} -s \\ 1 \end{pmatrix} \\
        &\quad\ + \frac{\left(2p\right)^{\tilde\absangparam}\Gamma(1-2\tilde\absangparam)}{\Gamma(1-\tilde\absangparam-\tomega/p)} x^{+\tilde\absangparam}
        \begin{pmatrix} s \\ 1 \end{pmatrix}, 
    \end{aligned}
\end{equation}
where we have taken the leading form of transformation defined in Eq.~\eqref{eq:psi_tranformation}. 

For later convenience, we introduce $\cR_f$ and $\cR_h$ as the ratios of coefficients in asymptotic expansions \eqref{eq:f2h} and \eqref{eq:h2f}, respectively. Explicitly, they are given as follows:
\begin{align}
    \cR_f &\approx \frac{(2p)^{-2\tilde\ell+s}}{(2imM)^s}\frac{\Gamma(\frac{1-s}
    {2}+2\tilde\ell)\Gamma(\frac{1+s}{2}-\tilde\ell-\tomega/p)}{\Gamma(\frac{1+s}
    {2}-2\tilde\ell)\Gamma(\frac{1-s}{2}+\tilde\ell-\tomega/p)}, \label{eq:Rf_RN} \\
    \cR_h &= -(2b)^{2\tilde\ell}\frac{\Gamma(\tilde\ell)\Gamma(-2\tilde\ell)\Gamma(\frac12+\tilde\ell-2i\omega_h)}{\Gamma(2\tilde\ell)\Gamma(-\tilde\ell)\Gamma(\frac12-\tilde\ell-2i\omega_h)}.
\end{align}
To obtain $\cR_f$ above, we have used the approximation
\begin{equation}
    \beta_+ \approx
    \begin{cases}
        2mM, & s=+1, \\
        \frac{p^2}{\tilde\ell^2}\frac{(\tilde\ell+\kappa/p)(\tilde\ell-\kappa/p)}{2mM}, & s=-1,
    \end{cases}
\end{equation}
which discriminates the sign of $\angparam$.

\subsection{Improved analytic formulas of quasibound-state spectrum}

The poles of Green's function $\tbG(x,y;\omega)$ correspond to zeros of the Wronskian $\mathcal W(\psi_\ing,\psi_\de)$. At these specific frequencies, a solution satisfying both the specified boundary conditions at the event horizon and at the spatial infinity exists. Intuitively, such a fermion surrounding a Reissner-Nordstr\"{o}m black hole can potentially form an atomlike structure. However, its eigenfrequencies $\omega$ are intrinsically complex \cite{Finster:1998ak}. This complex nature signifies that the system is not strictly \textit{bound} like hydrogen but is instead a \textit{quasibound} state slowly leaking into the hole.

Without loss of generality, we assume $\text{Re}\omega>0$, as the opposite case follows by symmetry. According to asymptotics of approximate solutions in the overlapping region, the eigenfrequencies correspond to the roots of the algebraic equation
\begin{equation}\label{eq:matching}
    \cR_f = \cR_h.
\end{equation}
In general, the ratio $\cR_f$ becomes large for small $\pmass M$ while $\cR_h$ remains finite. A consistent matching is allowed when
\begin{equation}\label{eq:quantization}
    \frac{1-s}2 + \tilde\absangparam - \frac{\tomega}{p} = -n + \delta,
\end{equation}
and $\delta$ is a sufficiently small quantity. Substituting the ansatz into Eq.~\eqref{eq:matching} and keeping only the linear term of $\delta$ yields 
\begin{equation}\label{eq:delta_RN}
    \begin{aligned}
        \delta(\omega) \approx 
        \frac{(-1)^{n}(4ib\pmass M)^s(4bp)^{2\tilde\ell-s}\Gamma(\tilde\absangparam)\Gamma(-2\tilde\absangparam)}{n!\Gamma(-n-2\tilde\absangparam+s)\Gamma(2\tilde\absangparam)\Gamma(-\tilde\absangparam)} &\\
        \ \times 
        \frac{\Gamma(\frac{1+s}2-2\tilde\absangparam)\Gamma(\frac12+\tilde\absangparam-2i\omega_{h})}{\Gamma(\frac{1-s}{2}+2\tilde\absangparam)\Gamma(\frac12-\tilde\absangparam-2i\omega_{h})} &. 
    \end{aligned}
\end{equation}

Now Eq.~\eqref{eq:quantization} is an equation of eigenfrequency $\omega$, which can be solved perturbatively. With the expression of $p$ and $\tomega$ in Eqs.~\eqref{eq:def_p} and \eqref{eq:def_tomega}, an equivalent form of Eq.~\eqref{eq:quantization} can be obtained as
\begin{align}\label{eq:recursion}
    \omega^2=m^2-\left[\frac{M(2\omega^2-m^2)-\omega qQ}{n+\tilde{\ell}+(1-s)/2-\delta}\right]^2.
\end{align}
According to the power counting, the second term is smaller than the first one. Therefore, it is shown that $\omega\approx m$ at the leading order. Substituting back into the equation yields the next-to-leading-order correction. Repeating this procedure iteratively, we obtain the following analytic expression for the spectrum:
\begin{widetext}\begin{equation}\label{eq:spectrum_RN}
    \begin{aligned}
        \frac{\omega}{m} \approx&\ 1-\frac{(\pmass M-qQ)^2}{2 \bar{n}^2} + \frac{(\pmass M-qQ)^3}{2\bar{n}^3}\left(\frac{3mM-qQ}{\ell} - \frac{15mM-3qQ}{4\bar n}\right) \\
        &\ -i(4bmM)^{2\tilde\ell_0}(mM-qQ)^{2\tilde\absangparam_0+2-s}
        \left[1 + \frac{mM-qQ}{2\bar n}\left(\frac{9mM-3qQ}{\ell} + \frac{mM-qQ}{\bar n}\right)\right]\Gamma_{n\tilde\absangparam_0s},
    \end{aligned}
\end{equation}
where $\bar n= n+\absangparam+(1-s)/2$ and
\begin{equation}
    \Gamma_{n\tilde\absangparam_0s} \equiv \frac{(-1)^{n+\frac{1-s}2}\Gamma(1+\tilde\absangparam_0)\Gamma(-2\tilde\absangparam_0)\Gamma(\frac{1+s}2-2\tilde\absangparam_0)}{n!\bar n^{2\tilde\ell_0+3-s}\Gamma(-n-2\tilde\absangparam_0+s)\Gamma(2\tilde\absangparam_0)\Gamma(1-\tilde\absangparam_0)\Gamma(\frac{1-s}{2}+2\tilde\absangparam_0)}\frac{\Gamma(\frac12+\tilde\absangparam_0-2i\omega_{h0})}{\Gamma(\frac12-\tilde\absangparam_0-2i\omega_{h0})}. 
\end{equation}
\end{widetext}
Quantities with subscript 0 are evaluated at $\omega=\pmass$, i.e., $\tilde\absangparam_0=\tilde\absangparam(\omega = \pmass)$ and $\omega_{h0}=\omega_h(\omega = \pmass)$.

Compared with the earlier result \cite{Ternov:1980st}, this work improves the analytic expression by incorporating higher-order corrections of effective angular quantum number $\tilde\ell_0 - \absangparam \approx \cO(m^2M^2)$. For illustration, the binding energies $\cB \equiv \re\omega - m$ for several QBSs are listed in Table~\ref{tab:BoundEnergy}. Although present method does not lead to a significant improvement in numerical accuracy over all existing results, the correction of the angular quantum number gives rise to the fine-structure splitting among states of the same principal number $\bar n$. 

To further elucidate this fine-structure splitting, we note that the correction term in Eq.~\eqref{eq:spectrum_RN} depends explicitly on the orbital quantum number through the factor $(3mM-qQ)/\ell$. Such dependence originates from resolving all the $1/x$ terms in the approximate Eq.~\eqref{eq:radial_far}. Consequently, states sharing the same principal number $\bar n$ but differing in $\ell$ acquire unequal binding energies. This is closely analogous to the fine structure of the hydrogen atom, in which subleading corrections split levels that are degenerate at leading order. At the present order, however, the states with $\angparam=\pm\ell$ remain degenerate.

Besides, the numerical calculations further resolve a residual fine-structure splitting between the states with same $\bar n$ and $|\angparam|$. This additional splitting comes from the $\mathcal O(1/x^2)$ terms neglected in the approximate Eq.~\eqref{eq:radial_far}. For a Schwarzschild black hole, this effect was calculated in detail in Ref.~\cite{Dolan:2015eua, Jentschura:2014uwa} using Foldy-Wouthuysen transformation combined with perturbation theory, yielding
\begin{equation}
    \frac{\mathcal B_S-\mathcal B_{S,\text{hyd}}}{m} \approx \frac{(mM)^3}{\bar n^4}\left(\frac{15}{8}-\frac{3\bar n}{2\ell+1}-\frac{3\bar n}{2\ell-s}\right).
\end{equation}
In the limit $q=Q=0$, our result Eq.~\eqref{eq:spectrum_RN} reproduces the first two terms in this expression but does not capture the last term (see also our previous work \cite{Chen:2025enc}). The latter carries the explicit dependence of the sign of $\angparam$. Its absence therefore explains the residual degeneracy between these states at the present order.

\begin{table}
    \centering
    \renewcommand{\arraystretch}{1.3}
    \begin{ruledtabular}
    \begin{tabular}{ccccc}
    $\bar n$ & $m_\ell$ & $\cB_{\text{hyd}}$ & $\cB_{\text{ana}}$ & $\cB_{\text{num}}$ \\
    \colrule
    1                  & $+1$ & $-0.00040500$                  & $-0.00040783$ & $-0.00042816$ \\
    \colrule
    \multirow{3}{*}{2} & $+2$ & \multirow{3}{*}{$-0.00010125$} & $-0.00010143$ & $-0.00010203$ \\
                       & $+1$ &                              & $-0.00010077$ & $-0.00010503$ \\
                       & $-1$ &                              & $-0.00010077$ & $-0.00010273$ \\
    \colrule
    \multirow{5}{*}{3} & $+3$ & \multirow{5}{*}{$-0.00004500$} & $-0.00004503$ & $-0.00004513$ \\
                       & $+2$ &                              & $-0.00004497$ & $-0.00004531$ \\
                       & $-2$ &                              & $-0.00004497$ & $-0.00004520$ \\
                       & $+1$ &                              & $-0.00004774$ & $-0.00004620$ \\
                       & $-1$ &                              & $-0.00004774$ & $-0.00004522$ \\
    \end{tabular}
    \end{ruledtabular}
    \caption{Binding energy $\cB \equiv \re\omega - m$ of selected QBS with $q=Q=0.1$ and $m=0.1$. The mass of the black hole is set to unity. For comparison, we list simultaneously the hydrogenlike spectrum $\cB_{\text{hyd}}$, our analytic result $\cB_{\text{ana}}$ and the numerical results $\cB_{\text{num}}$ from continued fraction method.}
    \label{tab:BoundEnergy}
\end{table}

    \begin{figure*}
		\centering
		\includegraphics[width=\linewidth]{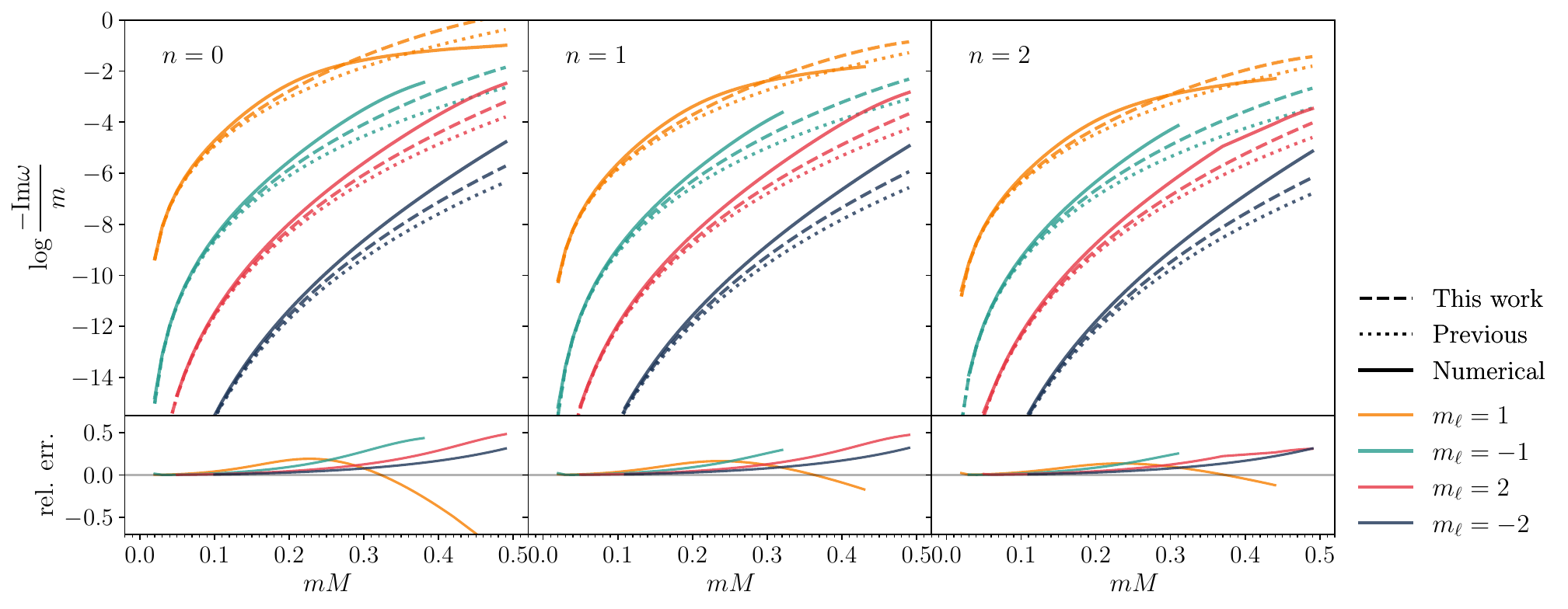}
		\caption{Comparison of analytic result of our work (dashed line) and in Ref.~\cite{Ternov:1980st} (dotted line) with numerical calculations (solid line) using matrix-valued continued fraction method for fixed $q=0.1$, $Q=0.1$ and $M=1$ and varying $m$. The upper panel illustrates the variation of results with parameter $m$, whereas the lower panel depicts the relative error defined in Eq.~\eqref{eq:RelErr}. The left, middle and right panels correspond to different overtone numbers and the angular quanta $\angparam$ is discriminated by different colors.}
		\label{fig:comparison_along_m}
    \end{figure*}

Besides the energy levels, the imaginary part of the eigenfrequency, i.e., the decay width, is more crucial for the fermion-BH system. Since the decay width determines the lifetime of the QBS, the systems can be long-lived and may be of phenomenological interest for light fermions \cite{Berti:2015itd, Coutant:2018bom, Dokuchaev:2014vda}. For comparison, we plot simultaneously our analytic result, leading-order approximation previously provided in Ref.~\cite{Ternov:1980st} and the numerical result by varying the mass of field $m$ in Fig.~\ref{fig:comparison_along_m}. We examine the relative error between our analytic result ($\im\omega_{\text{ana}}$) and the numerical result ($\im\omega_{\text{num}}$) as
\begin{equation}\label{eq:RelErr}
    \text{rel. err.} \equiv \frac{\im\omega_{\text{ana}} - \im\omega_{\text{num}}}{\im\omega_{\text{num}}}.
\end{equation} 
The plot is truncated at $mM=1/2$ for the case in which the wavelength of the field is comparable with the radius of the hole. The comparison shows that our improved analytic expression agrees well with the numerical analysis, especially for light fermions with $mM\leqslant0.1$. Compared with the previous result, Eq.~\eqref{eq:spectrum_RN} in general provides a better estimate of the decay width of this system. For the cases $\angparam\neq 1$, visible improvement has been achieved, especially for the mass range $mM=0.3$--$0.5$. The improvement comes from the incorporation of higher-order correction of angular quanta $\tilde\ell$ in our work. 

When $Q=0$, the expression reduces to the result for Dirac field around a Schwarzschild black hole, corroborating our previous calculation \cite{Chen:2025enc}. For a neutral field with $\pcharge=0$, the charged and uncharged black holes yield identical fine-structure spectra, differing only by a suppression factor $b^{2\tilde\absangparam_0}$ in the $\Gamma$-related term, which depends on the charge-to-mass ratio of black hole.


\subsection{Quasibound states outside an extremal Reissner-Nordstr\"{o}m black hole}

Having established the results for the nonextremal case, we proceed to the extremality limit $|Q|=M$ of the black hole. Since the approximate equation \eqref{eq:radial_near} becomes ill defined in this limit, it necessitates a separate treatment of eXRN black hole presented in this section.

The formalism reviewed in Sec.~\ref{sec:formalism} remains valid for the extremality limit with $\sqrt{\Delta(r)} = r-M$. Moreover, Eq.~\eqref{eq:radial_matrix} and the far-field approximation in \eqref{eq:radial_far} can also be safely extended to the extremal case $b=0$. The extremality only alters the dynamics in the \textit{near} region by changing Eq.~\eqref{eq:radial_near} into  
\begin{equation}
    \begin{aligned}
        \left(\partial_x + \frac{\tilde\absangparam}{x} - \frac{\hat\omega_h\hat{\kappa}_h}{\tilde\absangparam x^2}\right)\hat{\tilde\psi}^+_h &= \frac{i\hat\omega_h(1+i\hat\kappa_h/\tilde\absangparam)}{x^2}\hat{\tilde\psi}^-_h, \\
        \left(\partial_x - \frac{\tilde\absangparam}{x} + \frac{\hat\omega_h\hat{\kappa}_h}{\tilde\absangparam x^2}\right)\hat{\tilde\psi}^-_h &= \frac{i\hat\omega_h(1-i\hat\kappa_h/\tilde\absangparam)}{x^2}\hat{\tilde\psi}^+_h.
    \end{aligned}
\end{equation}
To derive the equations above, the same transformation defined in Eq.~\eqref{eq:psi_tranformation} has been employed. At the same time, the parameters are defined with a hat for the extremal case:
\begin{align}
    \hat\omega_h &= (\hat\omega\pm q)M, \\
    \hat{\kappa}_h &= (2\hat\omega\pm q)M . 
\end{align}
The sign $\pm$ in front of $q$ depends on the charge of the black hole. 

Such equations are solved also by Whittaker functions. Explicitly, we take
\begin{align}
    \hat{\tilde\psi}_{h1} &\approx \begin{pmatrix}
        (\tilde\ell+i\hat\kappa_h)
        W_{i\hat\kappa_h,\tilde\absangparam-\frac12}(-\frac{2i\hat\omega_h}{x}) \\
        (\tilde\ell-i\hat\kappa_h)
        W_{i\hat\kappa_h,\tilde\absangparam+\frac12}(-\frac{2i\hat\omega_h}{x})
    \end{pmatrix}, \\
    \hat{\tilde\psi}_{h2} &\approx \begin{pmatrix}
        (\tilde\ell-i\hat\kappa_h)
        W_{-i\hat\kappa_h,\tilde\absangparam-\frac12}(\frac{2i\hat\omega_h}{x}) \\
        (\tilde\ell+i\hat\kappa_h)
        W_{-i\hat\kappa_h,\tilde\absangparam+\frac12}(\frac{2i\hat\omega_h}{x})
    \end{pmatrix}. 
\end{align}
It is straightforward to verify that $\hat{\tilde\psi}_{h1} = \hat{\tilde\psi}_{h2}^*$. While the condition \eqref{eq:qbs_condition} holds for $|Q|=M$, $\hat{\tilde\psi}_{h1}$ is ingoing at the horizon. In the overlapping region, $\hat{\tilde\psi}_{h1}$ is asymptotically expanded as
\begin{equation}
    \begin{aligned}
        \hat\psi_{h1} &\approx -\frac{(-2i\hat\omega_h)^{\tilde\absangparam}\Gamma(1-2\tilde\absangparam)}{\Gamma(-\tilde\absangparam-i\hat\kappa_h)}x^{-\tilde\ell}
        \begin{pmatrix} s \\ 1\end{pmatrix} \\
        &\quad\ + \frac{(-2i\hat\omega_h)^{-\tilde\absangparam}\Gamma(1+2\tilde\absangparam)}{\Gamma(\tilde\absangparam-i\hat\kappa_h)}x^{+\tilde\ell}\xi_+ 
        \begin{pmatrix} -s \\ 1\end{pmatrix}. 
    \end{aligned}
\end{equation}
Thereby, the ratio becomes
\begin{equation}
    \hat\cR_h = (2i\hat\omega_h)^{-2\tilde\ell}\frac{\Gamma(2\tilde\ell)\Gamma(-\tilde\ell-i\hat\kappa_h)}{\Gamma(-2\tilde\ell)\Gamma(\tilde\ell-i\hat\kappa_h)}.
\end{equation}

The eigenfrequencies are still determined by Eq.~\eqref{eq:matching} but with the updated definition of $\hat\cR_{h}$ on the left-hand side. Assuming the consistency condition Eq.~\eqref{eq:qbs_condition}, we arrive at the estimate
\begin{equation}
    \begin{aligned}
        \delta(\hat\omega) \approx (-)^{n+\tilde\absangparam+1}
        \frac{(4i\hat\omega_hmM)^s(4\hat\omega_hp)^{2\tilde\absangparam-s}\Gamma(-2\tilde\absangparam)}{n!\Gamma(-n-2\tilde\absangparam+s)\Gamma(2\tilde\absangparam)} &\\
        \ \times \frac{\Gamma(\frac{1+s}{2}-2\tilde\absangparam)\Gamma(1-\tilde\absangparam+i\hat\kappa_h)}{\Gamma(\frac{1-s}{2}+2\tilde\absangparam)\Gamma(1+\tilde\absangparam+i\hat\kappa_h)} &.
    \end{aligned}
\end{equation}
Hence, the recursion yields the result
\begin{widetext}\begin{equation}\label{eq:spectrum_eXRN}
    \frac{\hat\omega}{\pmass} \approx 1 - \frac{\hat\omega_h^2}{2\bar n^2} + \frac{\hat\omega_h^3M}{2\bar n^3}\left(\frac{3\pmass-\pcharge}{\absangparam} - \frac{15\pmass-3\pcharge}{4\bar n}\right)
    -i(4mM)^{2\tilde\absangparam_0}\hat\omega_h^{4\tilde\absangparam_0+2-s}
    \left[1+\frac{\hat\omega_h M}{2\bar n}\left(\frac{9m-3q}{\absangparam}+\frac{m-q}{\bar n}\right)\right]\hat\Gamma_{n\tilde\absangparam_0s},
\end{equation}
with the quantum-number-dependent factor
\begin{equation}
    \hat\Gamma_{n\tilde\absangparam_0s} = \frac{(-1)^{n+\tilde\absangparam_0+\frac{1-s}{2}}\Gamma(-2\tilde\absangparam_0)\Gamma(\frac{1+s}{2}-2\tilde\absangparam_0)\Gamma(\tilde\absangparam_0-i\hat\kappa_h)}{n!\bar n^{2\tilde\absangparam_0+3-s}\Gamma(-n-2\tilde\absangparam_0+s)\Gamma(2\tilde\absangparam_0)\Gamma(\frac{1-s}{2}+2\tilde\absangparam_0)\Gamma(-\tilde\absangparam_0-i\hat\kappa_h)}.
\end{equation}\end{widetext}

\begin{figure*}
	\centering
	\includegraphics[width=\linewidth]{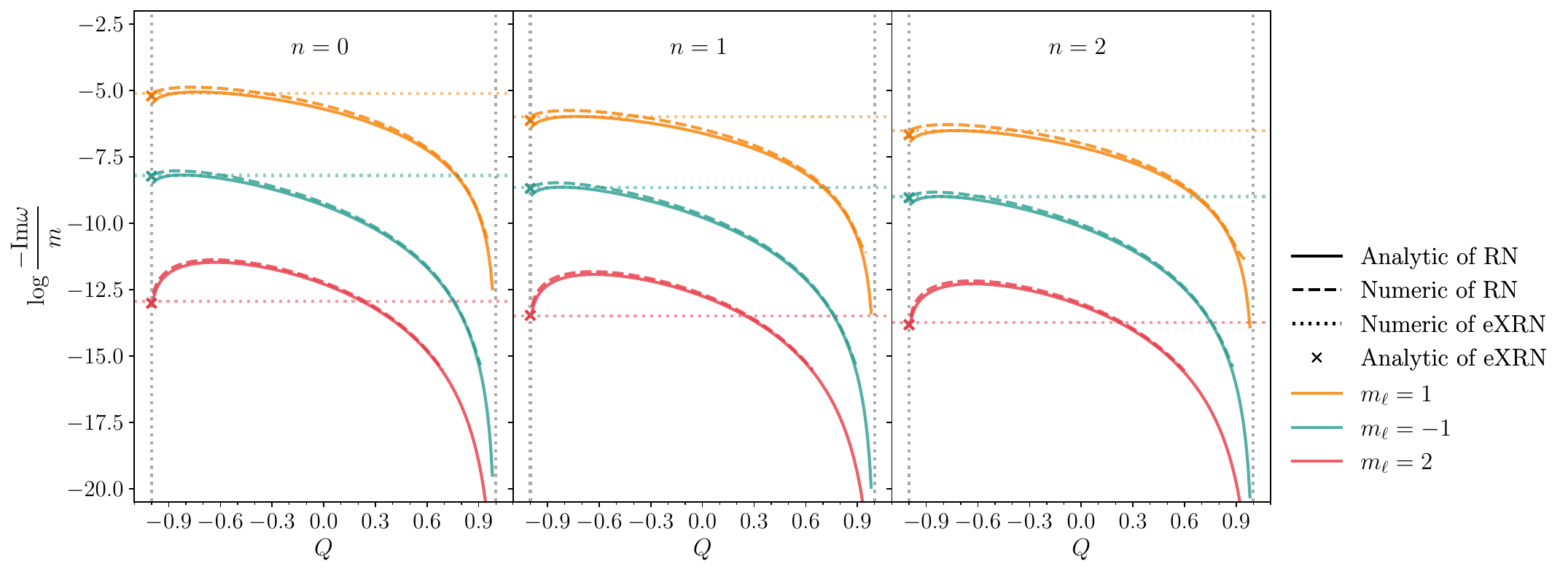}
	\caption{Comparison of analytic result (solid line) and numerical calculations (dashed line) using matrix-valued continued fraction method for fixed $q=0.1$, $m=0.1$ and $M=1$ and varying $Q$. The colorful dotted line indicates the numerical results obtained by shooting method for $Q=-M$, while the cross denotes the analytic estimation using Eq.~\eqref{eq:spectrum_eXRN}. The left, middle and right panels correspond to different overtone numbers and the angular quantum number $\angparam$ is discriminated by different colors.}
	\label{fig:comparison_along_Q}
\end{figure*}

Although the procedure does not work for the case of the eXRN BH, the final result in the last subsection Eq.~\eqref{eq:spectrum_RN} is well defined and finite in the limit $b\to0$. Compared with the spectrum of a nonextremal black hole, the hydrogenic and fine-structure terms are simply their smooth continuations. However, from the equality
\begin{equation}             
    \lim_{b\to0} b^{2\tilde\absangparam_0}\frac{\Gamma(\frac12+\tilde\absangparam_0-2i\omega_{h0})}{\Gamma(\frac12-\tilde\absangparam_0-2i\omega_{h0})} = (-i\hat\omega_h)^{2\tilde\absangparam_0},
\end{equation}
one can readily show that for a field around a black hole with fixed mass $M$, 
\begin{equation}
    \frac{\lim_{b\to 0}\omega(b)}{\hat\omega} \approx 1 + \cO(\hat\kappa_h).
\end{equation}
Equivalently speaking, the limiting value of Eq.~\eqref{eq:spectrum_RN} deviates from the result in this subsection Eq.~\eqref{eq:spectrum_eXRN} by a higher-order correction. Nevertheless, our numerical study reveals that this apparent deviation stems from an artifact of the approximation rather than being rooted in physics. This fact can be checked in Fig.~\ref{fig:comparison_along_Q}. The decay width for a RN black hole (dashed line) is calculated using matrix continued fraction method, while that for an eXRN black hole $Q=-M$ (dotted line) is obtained independently via the shooting method. The results rule out the spurious discontinuity at $Q=-M$. It is also found that Eq.~\eqref{eq:spectrum_eXRN} provides a more accurate prediction (cross markers) for the eXRN black hole that the limited value from our previous estimation Eq.~\eqref{eq:spectrum_RN}. This figure confirms also the vanishing of $\im\omega$ when the charge-to-mass ratio of the field and the black hole are reciprocal, i.e., $Q/M = m/q$.

\section{Branch-cut contribution and multistage late-time behavior}\label{sec:latetime}

\subsection{Branch-cut discontinuity and the late-time Green’s function}

A fermion captured by a black hole is generally not in one of the eigenstates discussed in the previous section. It is therefore necessary to study the time evolution of an arbitrary initial field configuration. In the Green-function representation, the pole contribution
describes the excitation of quasibound states, whereas the branch cuts attached to the mass thresholds encode the continuum response. At sufficiently late times, the isolated pole contribution and the large-arc contribution are exponentially suppressed. The remaining dominant signal is therefore controlled by the discontinuity across the branch cuts.

As shown in Fig.~\ref{fig:ComplexPath}, in this work branch cuts are placed at $\omega=\pm\pmass$, both extending parallel to the imaginary axis. With the fixed sign convention in Eq.~\eqref{eq:def_p}, this setup selects the branch for which the asymptotic mode at spatial infinity has the desired exponential behavior and therefore implements the appropriate decaying boundary condition along the deformed contour. It is important to keep this boundary condition explicit: throughout this section the quasibound Green function is defined by the decaying solution at spatial infinity rather than by the outgoing solution used in the usual QNM Green's function. Their contributions are then
\begin{equation}\label{eq:temp_Green}
    \bG_{\text{cut}} = \sum_{\xi=\pm}\int_0^{+\infty}\dd u\ e^{-i\xi mt-ut}\text{Disc}_\xi \tbG(\xi m-iu),
\end{equation}
where $\xi=\pm$ labels the discontinuity across the two branch cuts attached to $\omega=\pm\pmass$. Owing to the exponential factor $e^{-ut}$ in the integrand, the dominant contribution arises from the neighborhood of the branch points. Meanwhile, the key assumption \eqref{eq:assumption} holds and the approximate solutions in Eqs. \eqref{eq:psif} and \eqref{eq:psih} remain valid in the subsequent analysis. The late-time problem is thus reduced to evaluating the discontinuity of the scattering data near the two threshold branch points.

Since the four solutions defined in the last section are not all linearly independent, we introduce the connection formula that 
\begin{equation}\label{eq:conn_formulas}
    \psi_\de(x,\omega) = c_\ing(\omega)\psi_\ing(x,\omega) + c_\outg(\omega)\psi_\outg(x,\omega).
\end{equation}
With this connection formula, the retarded Green's function ($x<y$) can be expanded in the basis of horizon-asymptotic states, as follows:
\begin{equation}\label{eq:green_retard}
    \tbG = \frac{\psi_\ing(x)\bar\psi_\outg(y) + \mathcal S(\omega)\psi_\ing(x)\bar\psi_\ing(y)}{\mathcal W(\psi_\ing,\psi_\outg)},
\end{equation}
with $\mathcal S= c_\ing/c_\outg$ being the scattering factor. Such a factor can be determined by matching the known asymptotic expansion within the overlap region. Accounting for the transformation in Eq.~\eqref{eq:psi_tranformation} up to the leading order of $mM$, the matching yields 
\begin{equation}\label{eq:approxS_RN}
    \mathcal S = \frac{\cR_h^*+\cR_f}{\cR_f - \cR_h}.
\end{equation}
Since Eq.~\eqref{eq:radial_near} is equivalent to the description of a freely propagating and massless charged fermion, the first term of Eq.~\eqref{eq:green_retard} represents the direct propagation from $y$ to $x$ within the vicinity of the horizon. This direct term is analytic across the mass-threshold cuts and does not generate the late-time tail considered here. The nontrivial tail is encoded in the second term, which describes the wave backscattered by the
long-range far-field potential.

Consequently, the relevant discontinuity is induced by the jump in the scattering factor $\cS$. From the approximation \eqref{eq:approxS_RN}, it is shown that 
\begin{equation}
    \mathrm{Disc}_\pm\cS = \frac{(\cR_h+\cR_h^*)(\cR_f^+-\cR_f^-)}{(\cR_f^+ - \cR_h)(\cR_f^- - \cR_h)}.
\end{equation}
Here, we take $\cR_f^+\equiv \cR_f(p)$ and $\cR_f^-\equiv \cR_f(e^{i\pi }p)$ to lie on opposite sides of the branch cut. In obtaining this expression, $\cR_h$ is continuous across the cut; the discontinuity is generated by the replacement $p\to e^{i\pi}p$ in $\cR_f$. We will show below that the time dependence of the tail is mainly fixed by this jump $\cR^+_f-\cR^-_f$ in the far region, together with the denominator in the discontinuity formula above, while $\cR_h$ supplies the horizon scattering data.

\subsection{The corrected oscillatory power law in the intermediate late-time tails}

The late-time behavior is dominated by the neighborhoods near $\omega=\pm\pmass$. Although the size of these neighborhoods shrinks with increasing $t$, it is still sufficiently large in the \textit{intermediate late-time} regime that the effect of backscattering remains negligible. Accordingly, the inequality 
\begin{equation}
    \left|\frac{\kappa}{p}\right| \approx \frac{m(mM-qQ)}{\sqrt{u|2im-u|}} < \tilde\ell
\end{equation}
holds, leading to the range $m^3M^2<u<m$. Combining this with the estimate $t\sim 1/u$, the intermediate late-time window is identified as
\begin{equation}
    \frac1{\pmass} <  t < \frac1{\pmass^3M^2}.
\end{equation}

In this regime, one can drop the term $\kappa/p$ in the Gamma functions of $\cR^+_f=\cR_f(p)$ and use the reflection identity $\Gamma(z)\Gamma(1-z)=\pi/\sin(\pi z)$ in Eq.~\eqref{eq:Rf_RN} to arrive at the approximation
\begin{equation}
    \cR_f^+ \approx \frac{2\cos\tilde\ell_0\pi}{(2p)^{2\tilde\ell_0-s}(2imM)^s}
	\frac{\Gamma^2(\frac{1-s}{2}+2\tilde\ell_0)}{\Gamma^2(\frac{1-s}{2}+\tilde\ell_0)}.
\end{equation}
All other parameters are evaluated at $\omega=m$. It yields the discontinuity
\begin{equation}
    \mathrm{Disc}_+\cR_f \approx 2e^{-i\tilde\ell_0\pi}\cos(\tilde\ell_0\pi)\cR_f^+.
\end{equation}
Since $\cR_f\propto p^{-2\tilde\ell_0+s}$ and $|\cR_f/\cR_h| \gg 1$ in this regime, Eq.~\eqref{eq:approxS_RN} above gives $\mathrm{Disc}_+\cS \propto u^{\tilde\ell_0-s/2}$. The remaining Laplace integral then yields 
\begin{equation}
    \int^\infty_0\dd u\ e^{-ut}u^{\tilde\ell_0-s/2} = \frac{\Gamma(\tilde\ell_0+1-s/2)}{t^{\tilde\ell_0+1-s/2}}.
\end{equation}
Incorporating the near-horizon behavior $\cR_h$ as $\omega\to\pmass$, the contribution from this cut to the intermediate late-time evolution is estimated by
\begin{equation}\label{eq:int_Green}
\bG_{\text{cut},+} \underset{\text{int.}}{\sim} \frac{e^{-imt}}{t^{\tilde\ell_0+1-\frac{s}{2}}}
\end{equation}
in this intermediate large-$t$ limit. Here, we keep only the time dependent coefficient in
the Green’s function and drop the matrix describing the spinor structure. The exponent is controlled by the effective angular momentum $\tilde\ell_0$ rather than by the integer $\ell$, which shows a small departure from the familiar half-integer decay powers.


Following the same analysis, we can obtain the contribution from $\omega=-\pmass$ cut 
\begin{equation}
    \bG_{\text{cut},-} \underset{\text{int.}}{\sim} \frac{e^{imt}}{t^{\tilde\ell(-m)+1-\frac{s}{2}}}.
\end{equation}
In the case $q\ne 0$, the Coulomb coupling $(\pmass M-qQ)$ between the field and the black hole breaks the symmetry between the cut, namely $\tilde\ell(m)\neq \tilde\ell(-m)$. But the exponents differ only by a higher-order correction of $\cO(qQ)$ in the weak-coupling scenario. As a result, the two threshold contributions can have slightly different envelope powers when $qQ\neq0$ but this difference enters beyond the leading order and does not obscure the dominant \textit{oscillatory power-law} behavior. The exponent differs from those obtained for the scalar field \cite{Hod:1998ra, Koyama:2000hj,Koyama:2001ee,Koyama:2001qw} and the Proca field \cite{Konoplya:2006gq}, thereby providing a \textcolor{blue}{theoretical prediction} of spin discrimination in the late-time exponent. 
The oscillatory part appears to be rather universal, with its frequency determined by the mass of the Dirac field. In the massless limit $\pmass\to0$, the oscillatory feature is absent and the behavior reduces to a pure power law. 

For validation, we numerically solve the Eq.~\eqref{eq:radial} using the method of lines in the tortoise coordinate
\begin{equation}
    r_* = r+\frac{r^2_+}{r_+-r_-}\ln(r-r_+) - \frac{r_-^2}{r_+-r_-}\ln(r-r_-).
\end{equation}
The radial derivative is approximated via an eighth-order stencil and the time evolution is integrated using a fourth-order Runge-Kutta method. To incorporate the ingoing boundary at the horizon, we employ a sponge layer at $r_*=-2000M$ to allow only the ingoing wave to pass. The outer spatial boundary is placed sufficiently far away to prevent any numerical artifacts. During the simulation, no specific boundary condition is imposed at this boundary. This numerical setup is adequate for testing the intermediate tail because the leading intermediate
power law is determined by the local threshold behavior of the far-region potential and is insensitive to whether the asymptotic solution is ultimately chosen to be decaying or outgoing. The distinction between these two boundary prescriptions becomes essential only in the asymptotic regime discussed below.

\begin{figure}
	\centering
	\includegraphics[width=\linewidth]{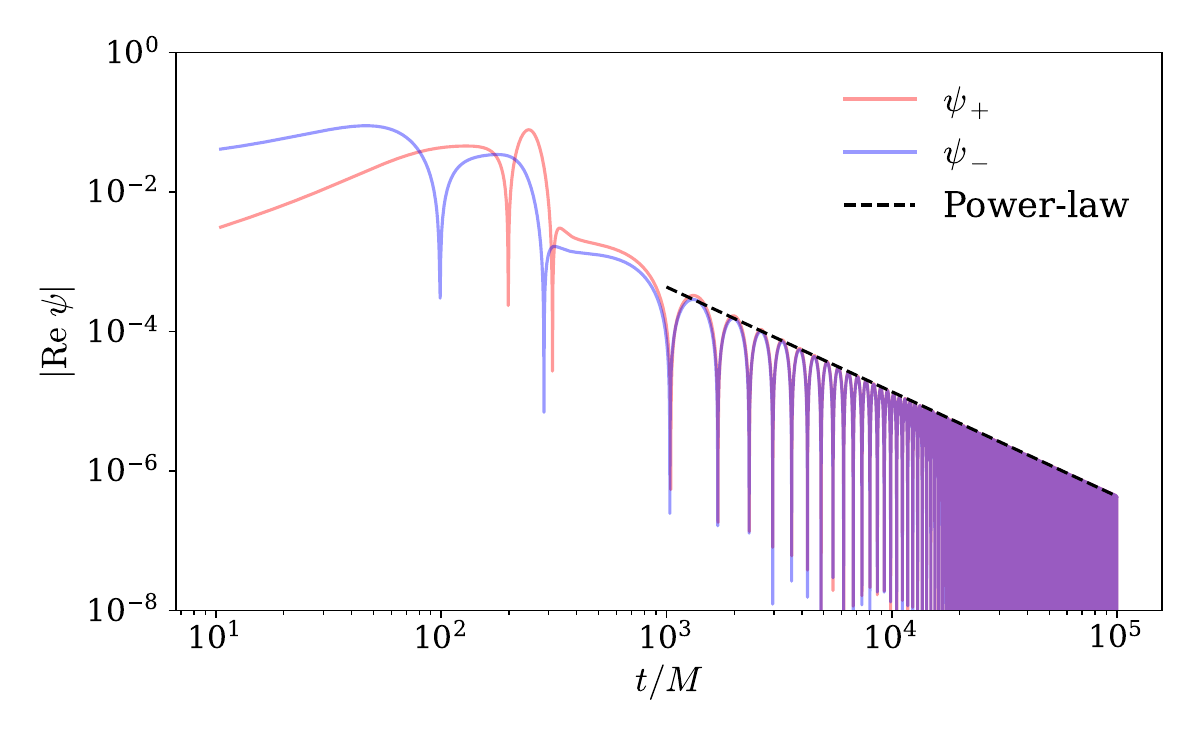}
	\caption{Numerical simulation of Eq.~\eqref{eq:radial} within $t=10$--$10^5M$. The wave was recorded at point $r_*=100M$, with parameters $\pmass = 0.005$, $M = 1.0$, $Q=0.01$, $q=0.3$ and $\angparam=+1$. The envelope is fitted to be $f(t)=14.91t^{-1.49}$, shown as black the dashed line in the figure.}
	\label{fig:IntermediateTail}
\end{figure}

Initially, a Gaussian wave packet is centered at $r_*=100M$ and evolved in time. To render the intermediate late-time regime observable, we set the parameters $mM=0.005$ and $qQ=0.003$. The numerical results are plotted in Fig.~\ref{fig:IntermediateTail}. In this regime, the wave envelope is fitted to be a power-law function $f(t) \propto t^{-1.49}$, which exhibits excellent agreement with our aforementioned analytical predictions. For $\angparam=+1$, the leading exponent predicted by Eq.~\eqref{eq:int_Green} is close to $3/2$, with a small correction, which is consistent with the fitted value.

\subsection{Far late-time tail: Stretched-exponential quasibound contribution}

Beyond the intermediate late-time scale, the behavior in the \textit{far late-time} regime 
\begin{equation}
     t\gg \frac1{\pmass^3M^2}    
\end{equation}
is more subtle and calls for careful treatment. The reason is that $\kappa/p$ gets larger and
\begin{equation}
    \frac{\Gamma(\frac{1+s}{2}-\ell-\frac{\kappa}{p})}{\Gamma(\frac{1-s}{2}+\ell-\frac{\kappa}{p})} \underset{u\to0}{\sim} \left(-\frac{\kappa}{p}\right)^{2\ell-s}
\end{equation}
holds when we explore the neighborhood much closer to the branch points.

In this case, it is shown that the asymptotic expansion holds near the positive-frequency branch point
\begin{equation}
\begin{aligned}
    \cR_f^+ \underset{u\to0}{\approx} &\ 
    \frac{2\sin(2\tilde\ell_0\pi)\Gamma^2(2\tilde\ell_0+\frac{1-s}{2})e^{-i\pi(2\tilde\ell_0-\frac12)}}{(2mM)^s(2\kappa_0)^{2\tilde\ell_0-s}} \\
    &\ \times\left[1+e^{-2i\tilde\ell_0\pi-\frac{2i\kappa_0\pi}{p}} + \text{(higher order)}\right]
\end{aligned}
\end{equation}
with $\kappa_0=\kappa(m)$. The first term at leading order in the brackets is continuous across the branch cut attached to $\omega=+m$. The second term at next-to-leading order contributes to the discontinuity.
The relevant integral contains the exponential factor $\exp[-ut-(1+i)\frac{\kappa_0\pi}{M\sqrt{mu}}]$ whose saddle lies at $u\sim t^{-2/3}$. Evaluating at the saddle point produces both the stretched-exponential suppression factor and the subdominant chirping phase, which yields
\begin{equation}\label{eq:Green_far_p}
    \bG_{\text{cut}, +} \underset{\text{far}}{\sim}  \frac{e^{-\eta t^{1/3}}}{t^{5/6}}e^{-i(mt + \frac{\eta}{\sqrt3}t^{1/3})}.
\end{equation}
The parameter $\eta=\frac{3\sqrt{3}}{m^{1/3}}\left(\frac{9\kappa_0\pi}{4M}\right)^{2/3}$ is fixed by the long-range threshold interaction through $\kappa_0$. By invoking the charge-conjugation symmetry for the conjugated branch cut, the saddle-point approximation gives also that
\begin{equation}\label{eq:Green_far_m}
    \bG_{\text{cut}, -} \underset{\text{far}}{\sim}  \frac{e^{-\eta t^{1/3}}}{t^{5/6}}e^{i(mt + \frac{\eta}{\sqrt3}t^{1/3})}.
\end{equation}

In the above, the oscillatory phase exhibits a chirping behavior similar to the existing analyses of the QNM sector, although the coefficient of the chirping correction is different. In particular, the late-time contribution from the QNM sector takes the form \cite{Jing:2004zb}
\begin{equation}\label{eq:Green_QNM}
    \bG_{\text{cut,QNM}}\sim \frac{1}{t^{5/6}}e^{i(mt + \frac{2\eta}{\sqrt3}t^{1/3})} + \text{(c.c.)}.
\end{equation}
The essential difference lies in the envelope. For the quasibound Green function, the asymptotic contribution contains an additional factor $e^{-\eta t^{1/3}}$ multiplying the universal $t^{-5/6}$ prefactor. Thus, the QBS sector does not yield a pure power-law tail and should be distinguished from the conventional massive-field asymptotic tail. These two behaviors arise from different boundary prescriptions at infinity for the same threshold branch cut: the usual $t^{-5/6}$ component is associated with the outgoing sector, whereas the expression above selects the decaying quasibound sector. Equation~\eqref{eq:Green_QNM} can in fact be recovered by carefully revisiting the preceding analysis with the outgoing prescription changing the threshold discontinuity entering the saddle exponent. 

The stretched-exponential factor originates from the accumulation of high-overtone QBS modes near the mass threshold. As $|\omega|\to m$, these modes become increasingly long-lived and spatially extended. Their spacing becomes small enough that the collective response is encoded in the nonanalyticity of branch points. The saddle-point approximation samples precisely this near-threshold accumulation region, converting it into the stretched-exponential suppression. 

\begin{figure}
	\centering
	\includegraphics[width=\linewidth]{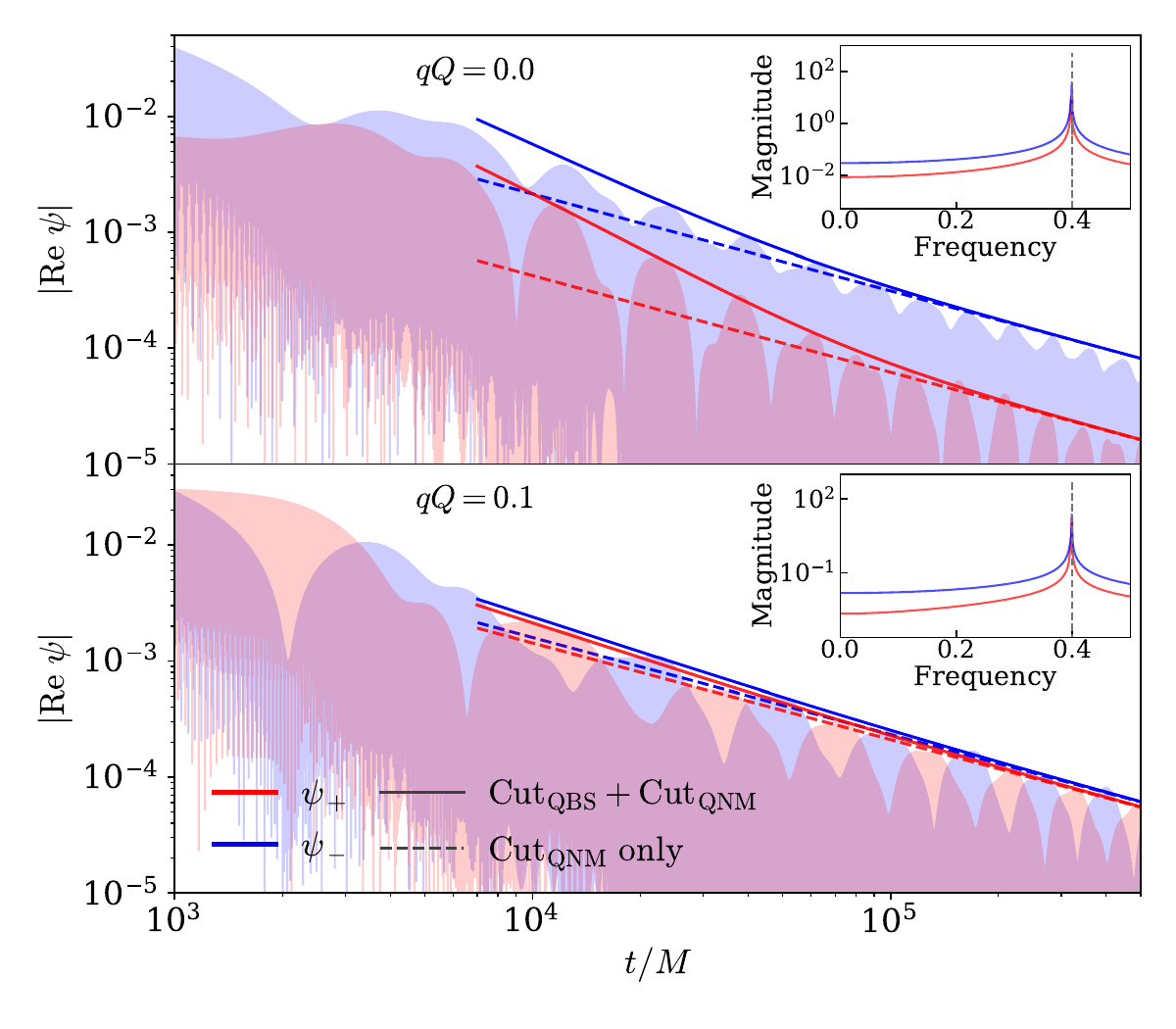}
	\caption{Comparison of the stretched-exponential and power-law fits for the late-time envelope of the Dirac field. In the simulation, we set $\pmass = 0.4$, $M = 1.0$, $Q=0.2$ and $\angparam=-1$. The top and bottom panel correspond $q=0.0$ and $q=0.5$, respectively. The insets show the Fourier spectra of the waveform for $tM>10^4$, over which the envelope is fitted to Eq.~\eqref{eq:EnvFit}.}
	\label{fig:latetimetail}
\end{figure}

We next test this asymptotic structure numerically. Figure~\ref{fig:latetimetail} illustrates the numerical results with $mM=0.4$ and $qQ=0.0$ as well as $qQ=0.1$. These parameter choices satisfy the condition Eq.~\eqref{eq:qbs_condition}, thereby ensuring the activation of QBS states. The relatively large value $mM = 0.4$ is chosen to make the asymptotic signal visible within a feasible evolution time. For this reason, the fitted coefficients shown in Table~\ref{tab:FitResult} should be regarded as a test of the functional form rather than as high-precision predictions of the weak-coupling formula. The late-time fitting was performed within the regime $tM>10^4$ where the characteristic frequency $\pmass$ is dominant (as evidenced by the Fourier spectrum in the inset). The envelope is fitted by the following function:
\begin{equation}\label{eq:EnvFit}
    f(t) = t^{-5/6}(\alpha_1 + \alpha_2e^{-\eta t^{1/3}}),
\end{equation}
which combines the QNM contribution with the QBS contributions. This mixed form is required because the finite-domain time evolution does not impose a purely quasibound boundary condition at the outer boundary. The initial wave packet can therefore project onto both the outgoing sector, represented by the coefficient $\alpha_1$, and the quasibound sector, represented by the coefficient $\alpha_2$.

The mixed fitting functions reflects the coexistence of two boundary-condition sectors in the numerical evolution, yielding qualitative agreement with the numerical data and separates the pure power-law component from the stretched-exponential QBS component. The smaller fitted value of $\eta$ in the charged case is also consistent with the analytic dependence of $\eta$ on the effective threshold coupling. We also note that for large mass parameters the performance of the analytical approximations slightly degrades, leading to discernible deviations from the numerical results.      

\begin{table}
    \centering
    \renewcommand{\arraystretch}{1.3}
    \begin{ruledtabular}
    \begin{tabular}{ccccccc}
           & $mM$ & $qQ$ & $\alpha_1$ & $\alpha_2$ & $\eta$ & $R^2$ \\
        \colrule
        $\psi_+$ & 0.4 & 0.0 & 0.910 & 53.847 & 0.124 & 0.999 \\
        $\psi_-$ & 0.4 & 0.0 & 4.597 & 111.698 & 0.124 & 0.997 \\
        $\psi_+$ & 0.4 & 0.1 & 3.081 & 6.818 & 0.070 & 0.991 \\
        $\psi_-$ & 0.4 & 0.1 & 3.463 & 7.871 & 0.070 & 0.998 \\
    \end{tabular}
    \end{ruledtabular}
    \caption{The fitted value of parameters in Eq.~\eqref{eq:EnvFit} and the $R$ squared.}
    \label{tab:FitResult}
\end{table}

\begin{figure}
	\centering
	\includegraphics[width=0.8\linewidth]{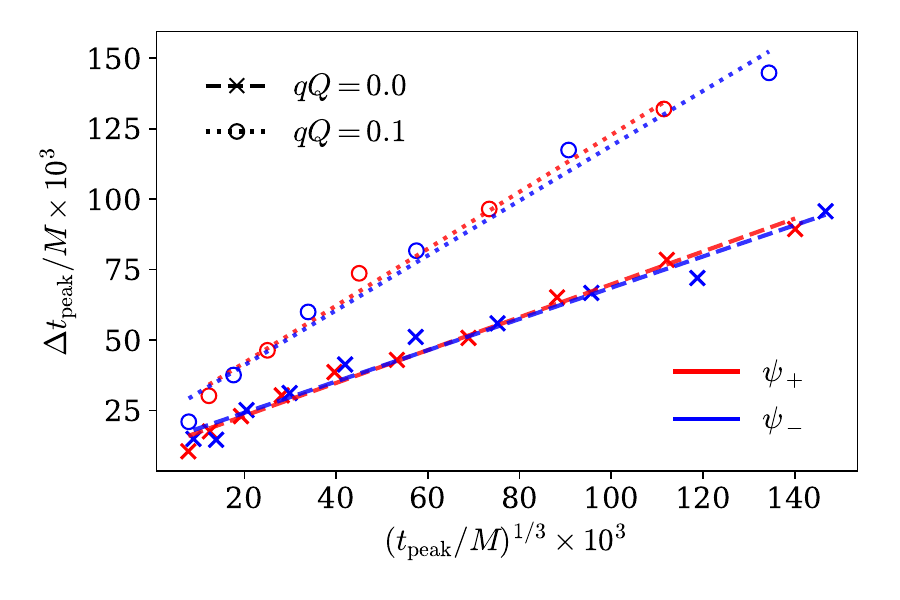}
	\caption{Time evolution of wavelength of the beat. The fitting confirms its linear dependence on $t^{1/3}$. }
	\label{fig:BeatingLength}
\end{figure}

At the same time, the numerical results shown in the figure also present a beat phenomenon in the far late-time regime. This modulation comes from the interference of the chirping frequencies of two components, as seen in Eqs.~\eqref{eq:Green_far_p}--\eqref{eq:Green_QNM}. In Fig.~\ref{fig:BeatingLength}, we explore the time evolution of the beat wavelength. It clearly demonstrates that its wavelength is scaled linearly with the term $t^{1/3}$.

For an eXRN black hole, simply replacing $\cR_h$ ($\cR_h^*$) by $-\hat{\cR}_h$ ($\cR_h^*$) suffices to replicate the primary conclusions for a nonextremal black hole. This is because the tail originates from the far-field effective potential and is largely insensitive to details near the horizon. The extremal geometry changes the threshold scattering amplitudes, but it does not change the type of branch-point nonanalyticity that controls the late-time
power law or the stretched-exponential factor derived below.

\section{Summary}\label{sec:summary}

In this work we have investigated the quasibound spectrum and late-time dynamics of a massive charged Dirac field in the exterior of a Reissner-Nordstr\"{o}m black hole. By formulating the radial Dirac equation as a matrix-valued first-order system, we constructed the Green's function with ingoing boundary conditions at the horizon and decaying boundary conditions at spatial infinity. This formulation provides a unified framework in which the pole contribution, associated with quasibound states, and the branch-cut contribution, responsible for late-time relaxation, can be analyzed consistently. 

The QBS spectrum is analytically estimated through the matrix matching scheme developed in Ref.~\cite{Chen:2025enc}. The inclusion of higher-order corrections to the effective angular momenta $\tilde\ell = \ell + \cO((mM)^2)$ gives rise to the fine-structure terms beyond the hydrogenic approximation. The accuracy of the decay width is also improved, especially for $mM$ values in the range 0.1--0.5.  The extremal RN case was treated separately, showing that its quasibound spectrum is smoothly connected to the extremal limit of the nonextremal result. Meanwhile, it is useful to adopt a separate analytic expression for extremal RN case for a better accuracy. 

On the other hand, the time-domain analysis shows that the relaxation of the fermionic cloud is dominated by the contribution from the branch cuts in the complex-$\omega$ plane. In the intermediate regime, they generate an oscillatory power-law tail whose exponent receives a correction from $\tilde\ell$. At asymptotically late times, the QBS boundary condition instead leads to a stretched-exponential contribution with a chirping phase. This behavior differs from the familiar QNM power-law tail and reflects the collective effect of highly excited, long-lived quasibound states. Our direct time-domain simulations reproduce both the intermediate power-law decay and the late-time signal containing the quasibound-state contribution, demonstrating the contribution of QBS at late times.  

\begin{acknowledgments}
    This work was supported in part by the National Science Foundation of China (NSFC) under Grants No. 12147103 (special fund to the center for quanta-to-cosmos theoretical physics), No. 11821505, and No. 12447105, the National Key Research and Development Program of China under Grant No. 2020YFC2201501, and the Strategic Priority Research Program of the Chinese Academy of Sciences under Grant No. XDB23030100.
\end{acknowledgments}

\section*{DATA AVAILABILITY}
There are no publicly available research data or software supporting this manuscript. Requests for further information or data should be sent to the authors.

\appendix

\section{VALIDITY RANGE OF APPROXIMATE SOLUTIONS}\label{app:valid_range}

In Sec.~\ref{sec:approxsols}, we seek the approximate solutions in the scenario Eq.~\eqref{eq:assumption}. The spatial infinity is an irregular singular point of Eq.~\eqref{eq:radial_matrix}. This fact is easily checked by expanding the operator at large-$x$ limit:
\begin{equation}
    \bigg(\partial_x\boldsymbol{I} - \sum\boldsymbol{W}^{(\infty)}_n x^n\bigg)\psi = 0.
\end{equation}
Here, $\boldsymbol{W}^{(\infty)}_n$ are the matrix-valued coefficients. To arrive at the approximate Eq.~\eqref{eq:radial_far}, we have kept the elements of order $\cO(mM)$ in $\boldsymbol{W}^{(\infty)}_0$ and that of order $\cO(mM/x)$ in $\boldsymbol{W}^{(\infty)}_1$. Simultaneously, elements of order $\cO(\ell/x^3)$ and $\cO(mM/x^2)$ are ignored in the subsequent terms in expansion. Hence, our approximation remains valid while the inequalities
\begin{equation}
    \min\left(mM,\frac{mM}{x}\right) > \max\left(\frac{mM}{x^2},\frac{\ell}{x^3}\right).
\end{equation}
It follows that Eq~\eqref{eq:radial_far} holds in the regime 
\begin{equation}
    x > \sqrt{\frac{\ell}{mM}},
\end{equation}
which is defined to be the \textit{far} region in the text.

The radial equation has a regular singular point at $x=b$ which corresponds to the location of horizon. The analysis near the horizon is performed in terms of a newly defined variable $x'=\sqrt{x^2-b^2}$, which asymptotically corresponds to the radius measured by spacelike proper distance. Expanding in $x'$, the radial equation is rewritten to be
\begin{equation}
    \begin{aligned}
        \left[x'\partial_x+\frac{i\omega_h}{x'}+\cO(mM)x'\right]\psi_+ &= \left[-\tilde\ell + \cO(mM)\right]\psi_-, \\
        \left[x'\partial_x-\frac{i\omega_h}{x'}-\cO(mM)x'\right]\psi_- &= \left[-\tilde\ell + \cO(mM)\right]\psi_+.
    \end{aligned}
\end{equation}
Since $\omega_h \sim \cO(mM)$, the approximate Eq.~\eqref{eq:radial_near} holds once
\begin{equation}
    \min\left(\frac{mM}{x'}, \tilde\ell\right) > \max(mM, mMx').
\end{equation}
Equivalently, our approximation is valid for 
\begin{equation}
    x < \frac{\ell}{mM},
\end{equation}
which is called \textit{near} region in the text.

\section{DETAILS OF NUMERICAL CALCULATION OF QUASIBOUND SPECTRA}

To complement our analysis, the radial Eq.~\eqref{eq:radial_matrix} was also investigated numerically using the matrix-valued continued fraction method \cite{Dolan:2015eua, Huang:2017nho}. First, the auxiliary radial functions $f_\pm$ are introduced via the ansatz
\begin{equation}\label{eq:harmonized_f}
    \begin{aligned}
        \psi^+ &= e^{-p x}(x+b)^{i\omega_h+i\tomega/p-\frac12}(x-b)^{-i\omega_h+\frac12}f^+, \\
        \psi^- &= e^{-p x}(x+b)^{i\omega_h+i\tomega/p}(x-b)^{-i\omega_h}f^-, \\
    \end{aligned}
\end{equation}
which serves to factor out the asymptotic behavior near singular points. Next, a Frobenius transformation
\begin{equation}
    u = \frac{x-b}{x+b}
\end{equation}
maps the domain of interest to a finite range $(0,1)$. 

These transformations recast the radial equation into a standard Frobenius form. Thus, the series $f(u)=\sum_n\balpha_n u^n$ constitutes a solution once the column-vector coefficient $\balpha_n$ satisfies a three-term recurrence relation
\begin{equation}
    \bU^{n-2}_2\balpha_{n-2} - 2\bU^{n-1}_1\balpha_{n-1} + \bU^{n}_0\balpha_{n} = 0.
\end{equation}
Here, the matrix-valued coefficients are given by
\begin{widetext}\begin{align}
	\bU^n_0 &= \begin{pmatrix} b(1-4i\omega_h + 2n) & 2b[\angparam-i\pmass M(1+b)] \\ 0 &  2bn \end{pmatrix}, \\
	\bU^n_1 &= \begin{pmatrix} b(2bp-\tomega/p+1+2n)-2i\omega M(1+b) + iqQ(2+b) & b[\angparam-i\pmass M(1-b)] \\ -b[\angparam+i\pmass M(1+b)] & b[2bp-\tomega/p-2i\omega M(1+b)+iqQ+2n]\end{pmatrix}, \\
	\bU^n_2 &= \begin{pmatrix} b(1-2\tomega/p+2n) - 2i[\omega M(1+b^2)-qQ] & 0 \\ -2b[\angparam+i\pmass M(1-b)] & -2b[\tomega/p+2i\omega M-iqQ - n] \end{pmatrix}.
\end{align}\end{widetext}
The subscript $n$ explicitly addresses its dependence on the index. For completeness and without loss of generality, we define $\balpha_{n} = \mathbf{0}$ for any $n<0$.

Since $f(0)=\balpha_0\neq0$, Eq.~\eqref{eq:harmonized_f} automatically satisfies the ingoing boundary condition at the event horizon. Conversely, the boundary condition at infinity is guaranteed only if the series $f(1) = \sum_n\balpha_n$ converges. The convergence thus imposes a constraint that determines the eigenfrequencies. The matrix continued fraction method provides an efficient way to solve this constraint \cite{Dolan:2015eua, Huang:2017nho}. Providing the step matrices $\bR_n$ satisfying 
\begin{equation}\label{eq:reccurence}
    \bfM_n = \bU_2^{n}\bR^{-1}_{n} - 2\bU_1^{n+1} + \bU_0^{n+2}\bR_{n+1} = 0,
\end{equation}
the eigenfrequency $\omega_{n_0}$ associated with overtone number $n_0$ is determined by the root of equation
\begin{equation}
    \det \bfM_{n_0}  = 0.
\end{equation}
Starting from $\bR_0 = 2\bU_0^{-1}(1)\bU_{1}(0)$, the above $\bR_{n_0}$ is computed forward by recurrence in Eq.~\eqref{eq:reccurence}. Meanwhile, $\bR_{n_0+1}$ is computed backward from a truncated term $\bR_{N}=\mathbf{1}$. The algebraic equation is solved by Nelder-Mead method with high-precision arithmetic implemented in \textsc{mpmath}.

\end{document}